\documentclass[reprint,aip,onecolumn]{revtex4-2}

\usepackage{graphicx}
\usepackage{nccmath}
\usepackage{epstopdf, epsfig}
\usepackage{amsmath, amssymb}
\usepackage{bm}
\usepackage{here}
\usepackage{graphicx,siunitx}
\usepackage{listings,color,courier,natbib}
\graphicspath{{}}
\usepackage[abs]{overpic}

\usepackage{color}
\definecolor{blue}{rgb}{0, 0.4470, 0.7410}
\definecolor{red}{rgb}{0.8500, 0.1250, 0.0480} 
\definecolor{orange}{rgb}{0.8500, 0.3250, 0.0980} 
\definecolor{yellow}{rgb}{0.9290, 0.6940, 0.1250}
\definecolor{purple}{rgb}{0.4940, 0.1840, 0.5560}
\definecolor{green}{rgb}{0.4660, 0.6740, 0.1880}

\draft 

\begin{document}

\title{Convolutional neural network and long short-term memory based reduced order surrogate for minimal turbulent channel flow}

\author{Taichi Nakamura}
\email[]{taichi.nakamura@kflab.jp}
\affiliation{Department of Mechanical Engineering, Keio University, Yokohama 223-8522, Japan}

\author{Kai Fukami}
\email[]{kfukami1@g.ucla.edu}
\affiliation{Department of Mechanical and Aerospace Engineering, University of California, Los Angeles, CA 90095, USA}
\affiliation{Department of Mechanical Engineering, Keio University, Yokohama 223-8522, Japan}

\author{Kazuto Hasegawa}
\email[]{kazuto-tess-1227@keio.jp}
\affiliation{Department of Mechanical Engineering, Keio University, Yokohama 223-8522, Japan}
\affiliation{Dipartimento di Scienze e Tecnologie Aerospaziali, Politecnico di Milano, Milano 20156, Italy}

\author{Yusuke Nabae}
\email[]{yusuke.nabae@kflab.jp}
\affiliation{Department of Mechanical Engineering, Keio University, Yokohama 223-8522, Japan}

\author{Koji Fukagata}
\email[]{fukagata@mech.keio.ac.jp}
\affiliation{Department of Mechanical Engineering, Keio University, Yokohama 223-8522, Japan}

\date{\today}

\begin{abstract}
We investigate the applicability of machine learning based reduced order model (ML-ROM) to three-dimensional complex flows. 
As an example, we consider a turbulent channel flow at the friction Reynolds number of $Re_{\tau}=110$ in a minimum domain which can maintain coherent structures of turbulence.
Training data set are prepared by direct numerical simulation (DNS).
The present ML-ROM is constructed by combining a three-dimensional convolutional neural network autoencoder (CNN-AE) and a long short-term memory (LSTM). 
The CNN-AE works to map high-dimensional flow fields into a low-dimensional latent space.
The LSTM is then utilized to predict a temporal evolution of the latent vectors obtained by the CNN-AE.
The combination of CNN-AE and LSTM can represent the spatio-temporal high-dimensional dynamics of flow fields by only integrating the temporal evolution of the low-dimensional latent dynamics.
The turbulent flow fields reproduced by the present ML-ROM show statistical agreement with the reference DNS data in time-ensemble sense, which can also be found through an orbit-based analysis.
Influences of the population of vortical structures contained in the domain and the time interval used for temporal prediction on the ML-ROM performance are also investigated.
The potential and limitation of the present ML-ROM for turbulence analysis are discussed at the end of our presentation.  

\end{abstract}


\maketitle 
\section{Introduction}

Reduced order modeling (ROM) methods have a crucial role to understand complex flow phenomena and to design control laws.
Among various methods, the proper orthogonal decomposition (POD)~\cite{Lumely1967} has vastly been utilized to extract dominant spatial coherent structures in flow fields~\cite{TBDRCMSGTU2017,THBSDBDY2019} and to construct Galerkin projection-based reduced order models~\cite{NAMTT2003,NPM2005}.
These efforts can be categorized as intrusive ROM, which fit modes into a lower order dynamical system.
It can retain much physical information of the original system because of its intrusiveness; however, there is a limitation on the applicability to complex flows where require an immense number of spatial modes~\cite{alfonsi2006,MPMF2019}, and on the stability in its numerical integration~\cite{ilak2008modeling}.
On the other hand, non-intrusive ROM, which extract governing physics from a vast amount of data sets, are able to handle the nonlinear complex flow phenomena efficiently~\cite{yu2019non}.
In particular, neural network based non-intrusive ROM have recently shown promising results.
San and Maulik~\cite{SM2018} introduced an extreme learning machine based ROM for quasi-stationary geophysical turbulent flows and showed its advantage in terms of stability against the POD-based ROM.
Otherwise, the great capabilities of neural network based ROM were demonstrated by considering two-dimensional Boussinesq equations at various Rayleigh numbers~\cite{pawar2019deep} and an inviscid transonic flow past an airfoil~\cite{renganathan2020machine}.
More recently, Maulik {\it et al.}~\cite{MFRFT2020} capitalized on a probabilistic neural network to obtain a temporal evolution of POD coefficients considering a shallow water equation and sea surface temperature.
Their proposed method can estimate the temporal dynamics while showing a confidence interval of its estimation. 
For more complex problems, Srinivasan {\it et al}~\cite{SGASV2019} applied a long short-term memory (LSTM)~\cite{HS1997} to the nine-equation ROM for turbulent shear flow and reported its ability to predict the temporal turbulent dynamics over coefficients. 
Moreover, Pawar {\it et al.}~\cite{pawar2020data} also utilized the LSTM so as to correct modal coefficients of the Galerkin projection.
They showed a great potential of their ROM when the governing equations are not sufficient to represent underlying physics.
In such a manner, the non-intrusive ROM aided by data-driven methods have acquired a citizenship in the fluid dynamics community as a powerful analysis tool.

In addition to the aforementioned studies, unsupervised machine learning based reduced order modeling with autoencoders (AEs)~\cite{HS2006} has also attracted an increasing attention due to its ability to account for nonlinearities in the low-dimensional mapping {{(in terms of vector dimension)}} via activation functions~\cite{BNK2020,BEF2019,BHT2020,FMF2020}.
In the pioneering work by Milano and Koumoutsakos~\cite{Milano2002}, who --- to the best of our knowledge --- first 
applied the AE to fluid flows, they investigated the similarity and difference between the POD and a multi-layer perceptron based AE considering the burgers equation and turbulent channel flow.
Recently, a convolutional neural network (CNN)-based AE has also widely been used thanks to the concept of filter sharing in CNN, which enables us to handle high-dimensional fluid data set efficiently~\cite{FFT2019b}.
Omata and Shirayama~\cite{omata2019} applied the CNN-AE and the POD to a wake of NACA0012 airfoil to reduce the dimension and examined the temporal behavior in the latent space.
From the perspective on interpretability of AE modes, Murata {\it et al.}~\cite{MFF2019} proposed a customized CNN-AE to extract and visualize the nonlinear AE modes by considering a two-dimensional periodic cylinder wake and its transient.
{
They clearly demonstrated that the CNN-AE is basically similar to POD but a single nonlinear AE mode contains multiple POD modes thanks to the nonlinear activation function.
}
The similar idea has also been extended by Fukami {\it et al.}~\cite{FNF2020} by utilizing a hierarchical CNN-AE to present ordered AE modes following the energy contributions aiming at more efficient and interpretable compression of turbulence.

More recently, Hasegawa {\it et al.}~\cite{HFMF2020a,HFMF2019} unified a CNN-AE and an LSTM to construct a machine learning based ROM (ML-ROM).
They examined its capability to unsteady flows around bluff bodies whose shape are randomly defined.
Their ML-ROM was also extended to the examination for Reynolds number dependence~\cite{HFMF2020b}.
These studies showed that the spatio-temporal high-dimensional dynamics of flow fields can be represented by only following the temporal evolution of the low-dimensional latent dynamics obtained by the CNN-AE.
However, their use of ML-ROM has been thus far limited to two-dimensional laminar flow fields.
Hence, our next interest is whether this ML-ROM can be applied to more complex flow field over a three-dimensional domain or not.

We here extend the ML-ROM constructed by CNN-AE and LSTM to a turbulent channel flow at $Re_{\tau}=110$ over a three-dimensional domain so that the applicability of the ML-ROM to three-dimensional complex flows can be investigated.
Our presentation is organized as follows.
Details of training data set and machine learning models are provided in section \ref{sec:methods}. 
Both qualitative and quantitative assessments on our ML-ROM for three-dimensional turbulence are shown in section \ref{sec:results}.
{We present a summary of our paper and give an outlook in section \ref{sec:con}.}

\section{Methods}
\label{sec:methods}

\subsection{Example: minimal turbulent channel flow at $Re_{\tau}=110$}

\begin{figure}
	\begin{center}
		\includegraphics[width=0.8\textwidth]{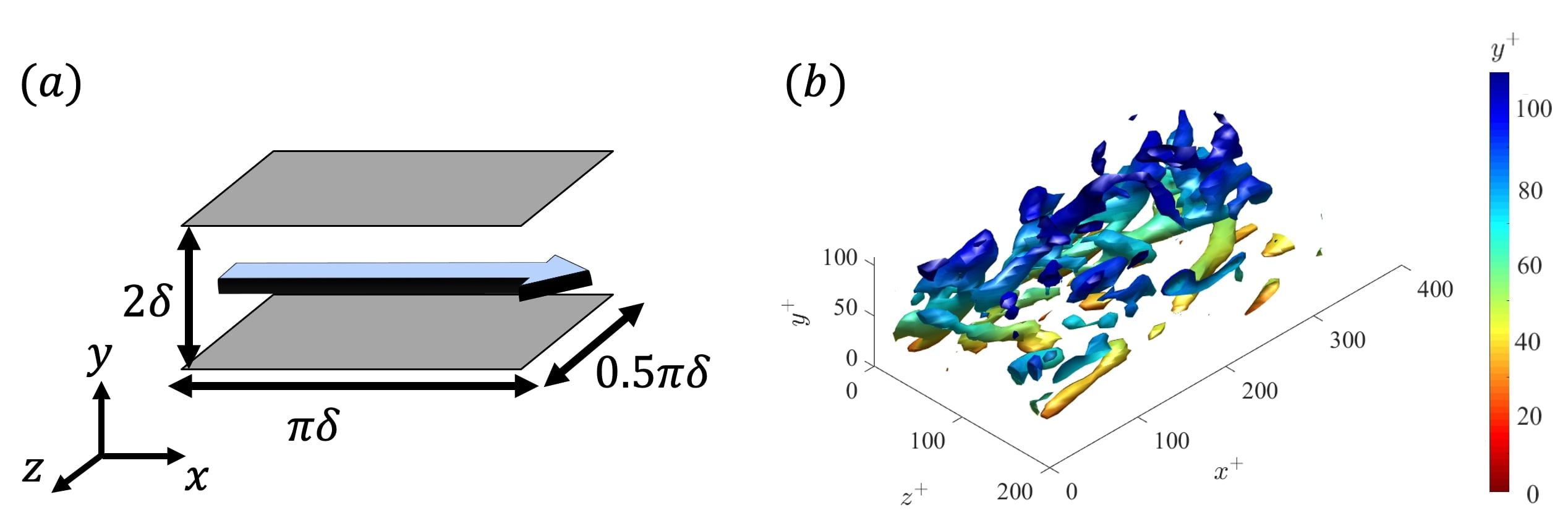}
		\caption{$(a)$ Computational domain of the present direct numerical simulation. $(b)$ Representative isosurface of the second invariant of velocity gradient tensor ($Q^+ = 0.01$).}
		\label{fig_DNS}
	\end{center}
\end{figure}

In the present paper, we consider a fully developed turbulent channel flow as an example of three-dimensional turbulence.
Training data set is prepared by a direct numerical simulation (DNS), which has been verified against the spectral DNS results by Iwamoto {\it et al.}~\cite{iwamoto2002reynolds}.
The governing equations are incompressible continuity and Navier--Stokes equations, i.e.,
\begin{align} 
&\bm{\nabla} \cdot {\bm u} = 0,\\
&{ \dfrac{\partial {\bm u}}{\partial t}  + \bm{\nabla} \cdot ({\bm u \bm u}) =  -\bm{\nabla} p  + \dfrac{1}{{Re}_\tau}\nabla^2 {\bm u}},
\end{align}
where $\displaystyle{{\bm u} = [u~v~w]^{\mathrm T}}$ represents the velocity vector with $u$, $v$ and $w$ being its components in the streamwise ($x$), wall-normal ($y$) and spanwise ($z$) directions. 
Here, $t$ is the time, and $p$ is the pressure.
{The quantities are non-dimensionalized with the fluid density $\rho^*$, the channel half-width $\delta^*$, and the friction velocity $u^*_\tau$, where the asterisk indicates dimensional quantities.}
The friction Reynolds number ${{Re}_\tau = u^*_\tau  \delta^*/\nu^*}$ and the computational domain size are carefully chosen so as to enable machine learning with our computational resource while maintaining coherent structures of turbulence{, referred to as a minimal turbulent channel flow}~\cite{jimenez1991minimal}.
The friction Reynolds number is set to ${Re}_\tau = 110$.
The size of the computational domain and the number of grid points here are $(L_{x}, L_{y}, L_{z}) = (\pi\delta, 2\delta, 0.5\pi\delta)$ and $(N_{x}, N_{y}, N_{z}) = (32, 64, 32)$, as illustrated in figure \ref{fig_DNS}$(a)$. 
The time step in the present DNS is $\Delta t_{\rm DNS}^+ {{=\Delta t_{\rm DNS}^\ast u_\tau^{\ast 2}/\nu^\ast}} =3.85\times 10^{-2}$,
where the subscript $+$ denotes the wall units.
{The governing equations (1) and (2) are spatially discretized with the energy-conserving fourth order finite difference scheme on the staggered grid~\cite{morinishi1998fully}.
The temporal integration is performed using the low-storage, third-order Runge-Kutta/Crank--Nicolson scheme~\cite{spalart1991spectral} with the high-order SMAC-like velocity-pressure coupling scheme~\cite{dukowicz1992approximate}. 
The pressure Poisson equation is solved with the fast Fourier transform in the $x$ and $z$ directions and the tridiagonal matrix algorithm in the $y$ direction.}
The computational grids in the $x$ and $z$ directions are uniform, and a non-uniform grid is utilized in the $y$ direction.
No-slip boundary condition is imposed on the walls and the periodic boundary condition is used in the $x$ and $z$ directions.

As for the data attributes used for machine learning, we consider three velocity components ${\bm q}=\{u,v,w\}$.
We use 10\,000 snapshots {(i.e., instantaneous velocity fields)} for training data as the baseline, although we will discuss the dependence of the CNN-AE performance on the amount of training data later.
These snapshots are sampled with a time interval $\Delta t^+ = 3.85$, which corresponds to 100 times that of DNS, i.e., $\Delta t^+=100\Delta t_{\rm DNS}^+$.
{For the training pipeline, we choose 70\% of the snapshots for the training of the present model, while remaining 30\% is used for the validation purpose.
The training for the LSTM is performed using the CNN-based latent vector generated using the DNS snapshots over $38\,500$ wall unit time corresponding to 10\,000 snapshots.
We also consider additional 2\,700 snapshots as the test data for the assessment in section~\ref{sec:results}.
Note that our test data are extracted from a different time period from that of the training data: the end of training snapshots and the beginning of the test data range are $231\,000$ wall unit time apart from each other.
Considering the fact that the temporal correlation vanishes on the order of $100$ wall unit time at this Reynolds number, the test data arrangement here can be regarded as far enough from the training data range, which can guarantee the generalizability of the present model.}
Note that we will discuss the dependence of the CNN-AE performance on the amount of training data later.
{In the present study, we do not apply any scalings (normalization and standardization) to both input and output attributes because all the attributes are more or less on the order of unity in wall units, although we can consider them to make the present model have the robustness for scaling with other flow fields~\cite{shanker1996effect}.}

\subsection{Convolutional neural network based autoencoder}

As mentioned above, the construction of our machine learning based reduced order model is inspired by Hasegawa {\it et al.}~\cite{HFMF2020a,HFMF2020b} who performed a temporal prediction of high-dimensional fluid flows in a low-dimensional latent space.
{Again, the role of CNN-AE in our ML-ROM framework is to map a high-dimensional flow field into a low-dimensional latent space.
Here, let us first introduce the principle of convolutional neural network (CNN)~\cite{LBBH1998}.}
The CNN has mainly been utilized in image processing and classification tasks.
Moreover, the use of CNN has also emerged in the fluid dynamics field because of the compatibility of filter sharing {(explained later)} idea to high-dimensional fluid data~\cite{FNKF2019,FFT2019a,FFT2019tsfp,CLKB2019,MFF2020,LTHL2020,KL2020,FHNMF2020,MFZF2020}.

\begin{figure}
	\begin{center}
		\includegraphics[width=1\textwidth]{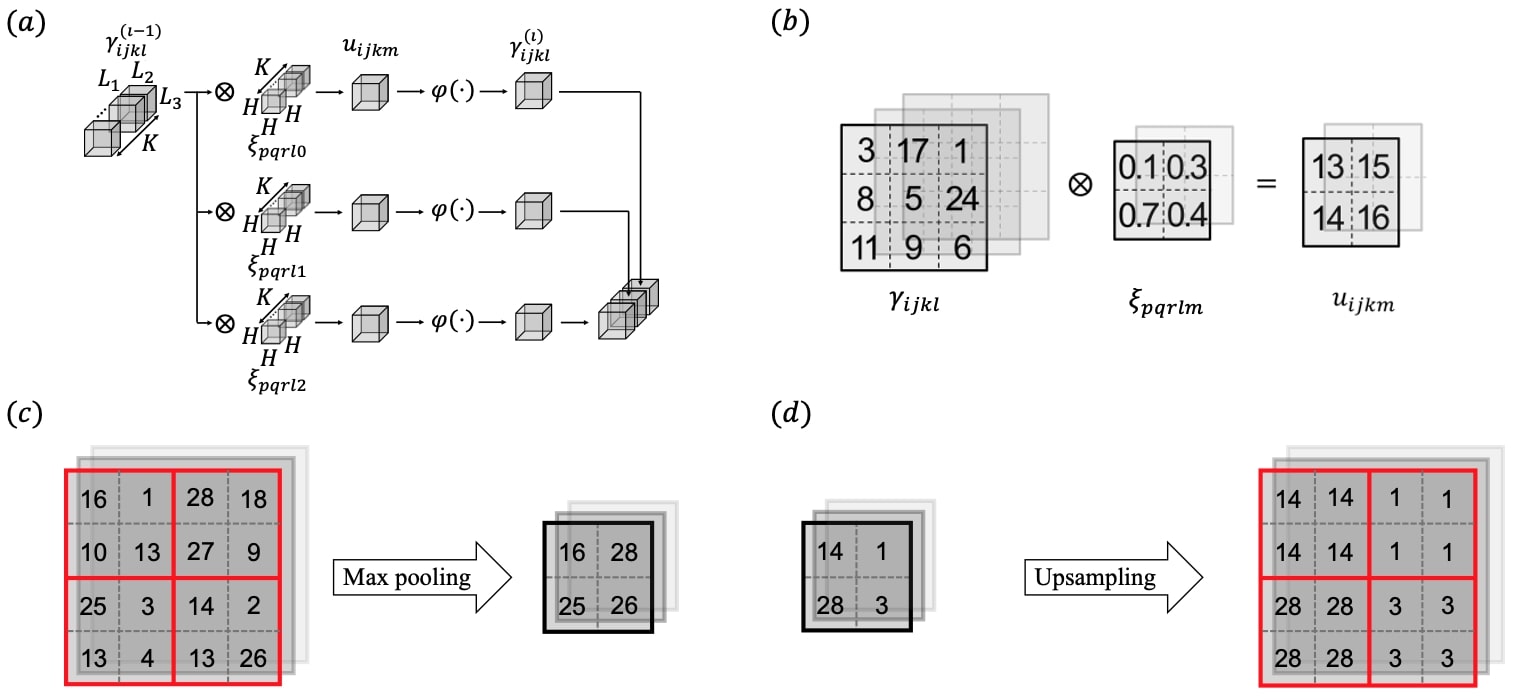}
		\caption{Internal procedure of convolutional neural network {(CNN)}. $(a)$ Three-dimensional convolution operation. {$(b)$ Filter operation.} $(c)$ Max pooling. $(d)$ Upsampling. Note that we use two-dimensional examples for $(b)$, $(c)$, and $(d)$ for clarity of illustration.}
		\label{fig_CNN}
	\end{center}
\end{figure}

Since we consider the channel flow over a three-dimensional domain, we use a three-dimensional CNN, which is distinct from Hasegawa {\it et al}.~\cite{HFMF2020a,HFMF2020b} who capitalized on a two-dimensional model.
As shown in figure~\ref{fig_CNN}$(a)$, the output {$\gamma^{({\iota})}$} at the layer $\iota$ {with the location $(i,j,k)$ in the $m$th filter}, can be generated by convolutional operation $\xi^{(\iota)}$ as illustrated in figure~\ref{fig_CNN}$(b)$. 
Mathematically speaking, the output {$\gamma^{(\iota)}$} can be formed as
\begin{eqnarray}
    {\gamma^{(\iota)}_{ijkm}} = {\varphi}\biggl({b_m^{(\iota)}}+\sum^{L-1}_{l=0}\sum^{{H}-1}_{p=0}\sum^{{H}-1}_{q=0}\sum^{H-1}_{r=0}\xi_{p{q}rlm}^{(\iota)} {\gamma}_{i+p{{-G}}\,j+{q{-G}}\,k+r-G,l}^{(\iota-1)}\biggr),
\end{eqnarray}
where {$H$ is the width and height of the filter}, {$G={\rm floor}(H/2)$,} $K$ is the number of channels in a convolution layer, $b_m$ is the bias, and $\varphi$ is an activation function. 
In the present paper, we utilize a ReLU function~\cite{NH2010}, which is known to be effective in avoiding the weight update issue in deep CNNs.
The aim of training process in the CNN can be regarded as obtaining optimized filter coefficients $\xi_{p{q}rlm}$ so as to acquire the desired output.  
{The concept of filters in CNN is known as filter sharing since the filter of size $H \times H$ is shared over the whole field 
within the same layer.}
In addition to the filter operation, a max pooling operation (figure~\ref{fig_CNN}$(c)$) and an upsampling operation (figure~\ref{fig_CNN}$(d)$) are respectively inserted for the encoder and decoder parts of autoencoder (AE), although details of AE will be provided later. 
The max pooling operation plays a role to reduce the dimension of input images while obtaining the robustness against rotation and translation of the images.
On the other hand, the upsampling operation in the decoder part copies the values of the lower-dimensional maps into a higher-dimensional image.

In the present study, the aforementioned CNN is utilized for the construction of autoencoder (AE).
The AE is one of unsupervised methods which use the same data ${\bm q}$ for the input and output in the training,
{
\begin{eqnarray}
    {\bm q}\approx \widehat{\bm q} = {\cal F}_C({\bm q};{\bm w}_C),
\end{eqnarray}
}
where ${\cal F}_C$ and ${\bm w}_C$ represent the
AE and the weights inside the AE, respectively.
{The output of the AE is denoted as $\widehat{\bm q}$.}
The AE generally has a compression procedure that comprises of an encoder ${\cal F}_e$ for dimension reduction and a decoder ${\cal F}_d$ for dimension extension.
If the output data of AE has a similar nature to the input $\bm q$, it implies that low-dimensional map of AE $\bm \eta$ can be regarded as representative information of high-dimensional original input or output such that 
\begin{eqnarray}
    {\bm \eta}= {\cal F}_e({\bm q}), ~~~{\widehat{\bm q}}= {\cal F}_d({\bm \eta}).
\end{eqnarray}
{As briefly mentioned in introduction, the low-dimensionalized variables arranged by the nonlinear AE have more energetic information than that with linear POD thanks to the use of nonlinear activation function~\cite{MFF2019,FNF2020}, although their
orthogonalities are not guaranteed~\cite{FHNMF2020}.}

\begin{figure}
	\begin{center}
		\includegraphics[width=0.85\textwidth]{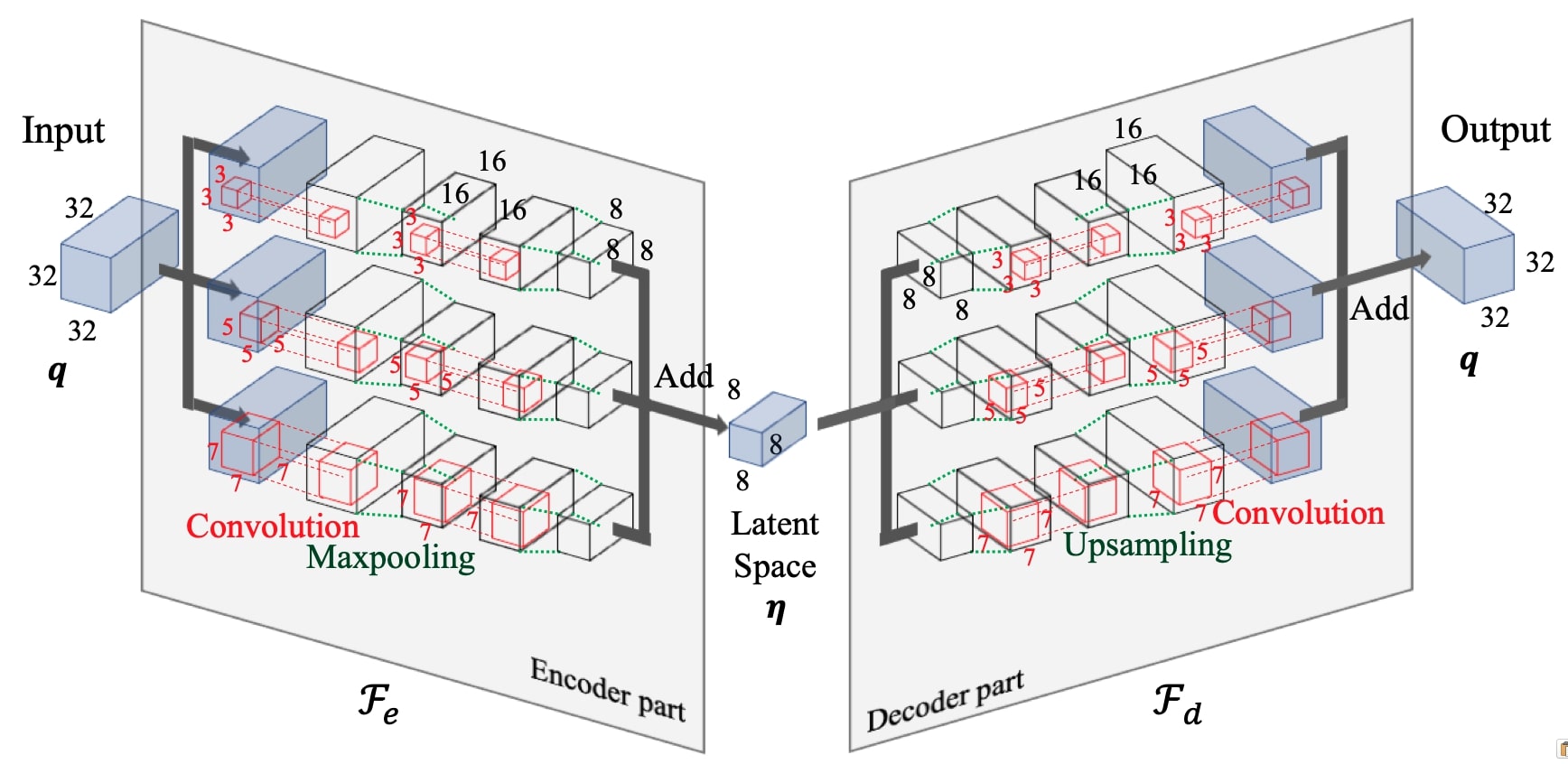}
		\caption{Three-dimensional multi-scale convolutional neural network based autoencoder {(CNN-AE)}.}
		\label{fig_MSCNNAE}
	\end{center}
\end{figure}

Analogous to our previous studies~\cite{HFMF2020a,HFMF2020b}, we use a multi-scale CNN-AE which contains three filter sizes as illustrated in figure~\ref{fig_MSCNNAE}.
One of significant advantages for the use of multi-scale (MS) manner is that the model is able to account for various scales included in complex fluid flow phenomena~\cite{FFT2019a}.
The sizes of filter are set to 3, 5, and 7 for each path in figure~\ref{fig_MSCNNAE}.
The details for the size of input/output data and low-dimensional latent space will be provided later.
The weights inside the multi-scale CNN-AE are obtained through an iterative minimization.
As stated above, the CNN-AE is trained to output the same data as the input ${\bm q}$,
\begin{eqnarray}
    {\bm w}_C = {\rm argmin}_{{\bm w}_C}||{\bm q}-{\cal F}_C({\bm q}; {\bm w}_C)||_2. \label{eq:CNNL2}
\end{eqnarray}
Note again that we use three velocity components ${\bm q}=\{u,v,w\}$ as the input and output attributes.
We use the $L_2$ error as the loss function, as shown in equation (\ref{eq:CNNL2}).

\subsection{Long short-term memory}

A long short-term memory~\cite{HS1997} is one of recurrent neural networks (RNNs) which can retain a temporal sequential history of time-series data into the model.
Due to its characteristics, the LSTM has often been utilized in speech recognition tasks~\cite{graves2013hybrid}.
In the present ROM, the LSTM is applied to the temporal prediction of low-dimensional latent space obtained by the CNN-AE introduced above.

\begin{figure}
	\begin{center}
		\includegraphics[width=0.90\textwidth]{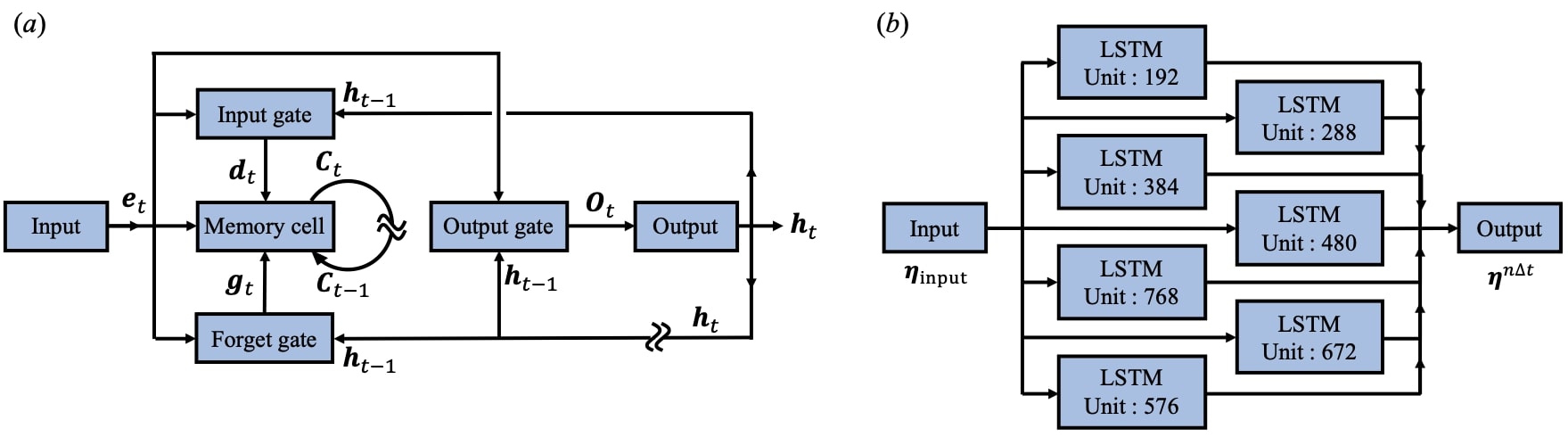}
		\caption{$(a)$ Internal procedures of long short-term memory {(LSTM)}. 
		{$(b)$ 
		Parallel
		use of 
		LSTM models having
		different number of units, adopted
		in the present study. 
		The output is obtained as the summation of these LSTM models.
		}}
		\label{fig_LSTM}
	\end{center}
\end{figure}

An LSTM layer is comprised of a cell, an input gate $d$, an output gate $o$, and a forget gate $g$.
Internal procedures in the LSTM layer is illustrated in figure~\ref{fig_LSTM}$(a)$, which can be mathematically expressed as  
\begin{align}
g_t&=\sigma(W_g\cdot [h_{t-1}, e_t]+\beta_g),\\
d_t&=\sigma(W_d\cdot [h_{t-1}, e_t]+\beta_d),\\
\widetilde{C}_{t}&=\tanh({W_c\cdot [h_{t-1}, e_t]+\beta_c}),\\
C_t&=g_t\times C_{t-1}+d_t\times\widetilde{C}_t,\\
o_t&=\sigma(W_o\cdot [h_{t-1}, e_t]+\beta_o),\\
{h}_t&=o_t\times\tanh(\it{C_t}),
\end{align}
where $C$ is the cell state, $h$ is the cell output, and $e$ denotes the cell input,
$W$ and $\beta$ represent the weights and the bias for each gate denoted by its subscript; the subscripts $t$ and $t-1$ for $C$, $e$, and $h$ represent the time indices, and $\sigma$ is the sigmoid function.
The LSTM layer can deal with a time-series problem by keeping the previous input information in the cell state $C$.

In the present study, we propose a parallel use of LSTM as shown in figure~\ref{fig_LSTM}$(b)$.
The parallel setting with various numbers of units here enables us to handle time-series data governed by complex nature, e.g., turbulence.
{Note that we can expect the use of sophisticated optimizations including hyperopt~\cite{BYC2013} and Bayesian optimization~\cite{MMLMBL2019} in order to improve the accuracy of the model, although we do not consider here.}
We have confirmed that the parallel LSTM outperforms the standard LSTM in our preliminary investigation.
The weights inside the LSTM are obtained through an iterative minimization similarly that for the CNN-AE.
{The present LSTM is trained to output a future state of the latent vector ${\bm \eta}^{n\Delta t}$ from the information at the five previous time steps ${\bm \eta}_{\rm input} = \{{\bm \eta}^{(n-1)\Delta t},{\bm \eta}^{(n-2)\Delta t},{\bm \eta}^{(n-3)\Delta t},{\bm \eta}^{(n-4)\Delta t},{\bm \eta}^{(n-5)\Delta t}\}$ such that 
\begin{eqnarray}
    {\bm w}_L = {\rm argmin}_{{\bm w}_L}||{\bm \eta}^{n\Delta t}-{\cal F}_L({\bm \eta}_{\rm input}; {\bm w}_L)||_2,
\end{eqnarray}
}
where ${\bm w}_L$ denotes the weights in the LSTM, and ${\cal F}_L$ indicates the LSTM model.
{For the test assessment, the latent variables for the previous five time step are generated from DNS data that are not used during the training for the CNN-AE.}
Our preliminary test has shown that the results are not sensitive to the number of time steps used for the input.

\subsection{CNN-LSTM based reduced order model}
\label{sec:CNN-LSTM}

\begin{figure}
	\begin{center}
		\includegraphics[width=0.9\textwidth]{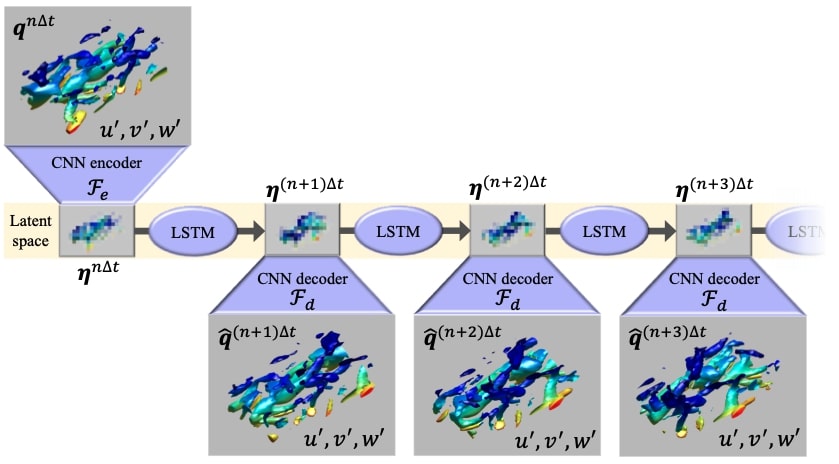}
		\caption{{Schematic of the present ML-ROM for turbulent channel flow.}}
		\label{fig_MLROM}
	\end{center}
\end{figure}

As illustrated in figure~\ref{fig_MLROM}, the proposed machine-learning based reduced order model (ML-ROM) consists of the CNN-AE model and the LSTM model.
{Five consecutive flow fields obtained by the DNS are fed into the trained single CNN-AE to obtain their low-dimensional representations as the five consecutive latent vectors used as the initial condition for the LSTM model.}
The LSTM model then predicts the latent vector at the next time step.
By using the previous output as the input, the LSTM model recursively predicts the temporal evolution of the encoded fields.
The temporal evolution of the flow field in the high dimensional space can be reconstructed by the trained CNN decoder with the predicted latent fields by the LSTM.

\section{Results and discussion}
\label{sec:results}
\subsection{Spatial order reduction via CNN}

\begin{table}[b]
\caption{Latent vector size of the CNN-AE models.}
\begin{center}
\begin{tabular}{cccc}
\hline\hline
    Case & $\eta_c$ & Latent vector\\ \hline
    Large  & 0.125 & $(16,16,16,3)$     \\ 
    Medium  & $1.56\times10^{-2}$  & $(8,8,8,3)$  \\ 
    Small  & $1.95\times10^{-3}$ & $(4,4,4,3)$ \\ 
    Extra Small & $2.44\times10^{-4}$ & $(2,2,2,3)$ \\ \hline\hline
    \end{tabular}
    \end{center}
    \label{table1}
\end{table}

Let us first examine the mapping ability of the CNN-AE part only, focusing on the size of latent space.
Four CNN-AE cases are considered, as summarized in Table~\ref{table1}.
The ratio of the size of latent space to that of high-dimensional space $\eta_c$ is defined as
\begin{equation}
\eta_c = \frac{N_x^{\#}\times N_y^{\#}\times N_z^{\#}\times N_\Phi}{N_x\times (N_y/2)\times N_z\times N_\Phi},
\end{equation}
where $N_\Phi$ is the number of physical attributes, and $(\cdot)^{\#}$ represents the size of latent space.
Note that we use a half domain of turbulent channel flow by considering the statistical symmetry in the $y$ direction.
The Adam algorithm~\cite{kingma2014} is applied as the optimizer for weight updating, and early stopping criteria~\cite{prechelt1998} is used to train the models and avoid overfitting.

\begin{figure}
	\begin{center}
		\includegraphics[width=0.75\textwidth]{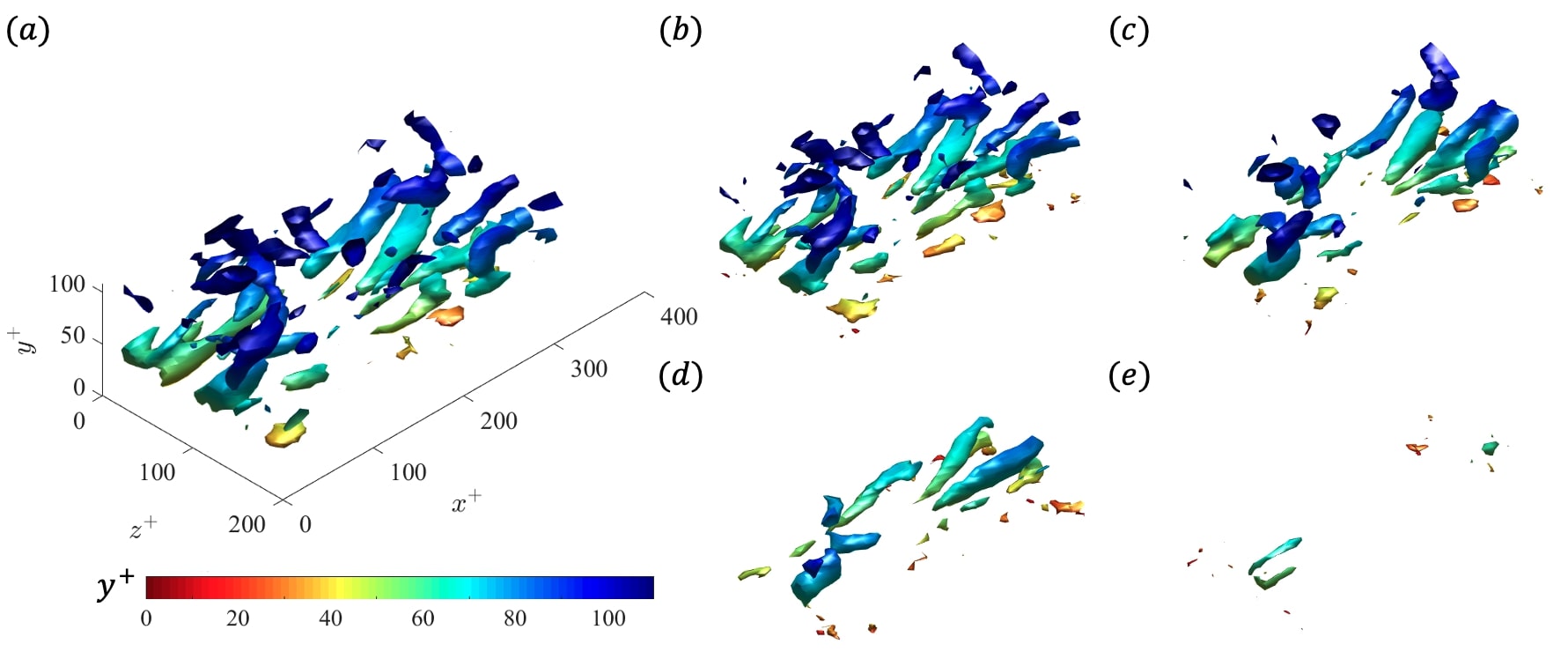}
		\caption{Isosurfaces of the second invariant of velocity gradient tensor ($Q^+=0.01$). $(a)$ Reference DNS. CNN-AE with $(b)$ large, $(c)$ medium, $(d)$ small, and $(e)$ extra small latent spaces.}
		\label{fig_QCNN}
	\end{center}
\end{figure}

The reconstructed instantaneous flow fields are visualized using the second invariant of velocity gradient tensor, as shown in figure~\ref{fig_QCNN}.
As can be found in figure~\ref{fig_QCNN}, not only large-scale vortex structures but fine structures can also be reconstructed by the large and medium models, which implies that the machine learning models can successfully low-dimensionalize the flow field.
In addition, the small model can also represent large-scale vortical structures, although the small-scale structure is not reconstructed.
This is likely because the information of small-scale structures is lost by max pooling and upsampling in the CNN-AE.
This trend due to the overcompression can also be found in the extra small model.
Summarizing the observation in figure~\ref{fig_QCNN}, the number of AE modes that can be handled by the small ($\in {\mathbb{R}}^{192}$) and extra small ($\in {\mathbb{R}}^{24}$) models is insufficient to reconstruct turbulent channel flow, even if the CNN-AE has a nonlinear activation function.

\begin{figure}
	\begin{center}
		\includegraphics[width=0.75\textwidth]{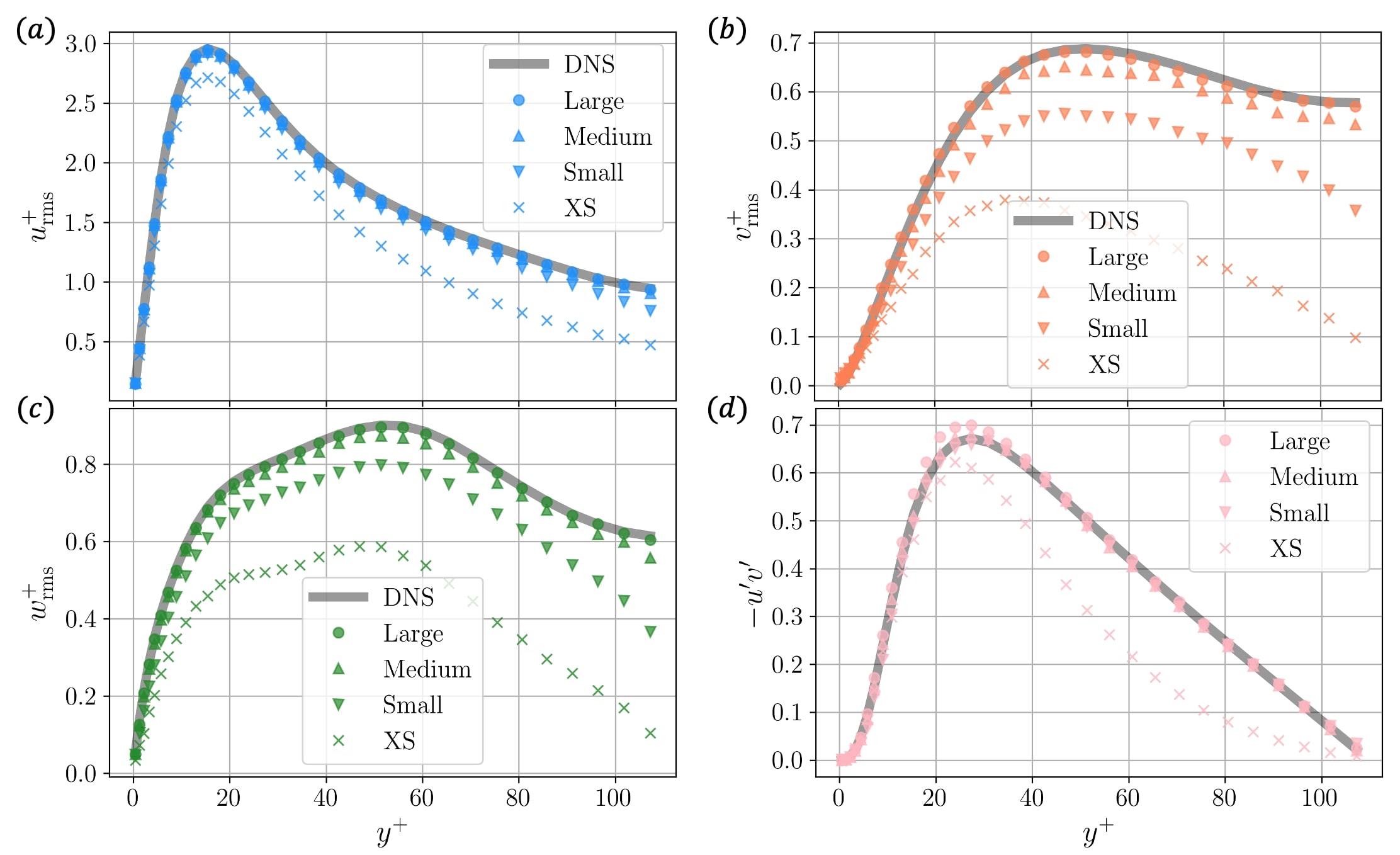}
		\caption{Statistics of the flow field reconstructed by CNN-AE models. $(a)$ $u^{+}_{\rm rms}$. $(b)$ $v^{+}_{\rm rms}$. $(c)$ $w^{+}_{\rm rms}$. $(d)$ $-\overline{u^\prime v^\prime}$. XS represents the extra small model.}
		\label{fig_RMSCNN}
	\end{center}
\end{figure}

We then assess the ability of the CNN-AE with statistical assessments, as summarized in figure~\ref{fig_RMSCNN}.
For both the root-mean-squared (RMS) values of all velocity components and the Reynolds shear stress, the curves reconstructed by the large and medium models are in reasonable agreement with the reference DNS data.
In contrast, the curve obtained by the small model is in reasonable agreement with the DNS for $u^+_{\rm rms}$ and $-u^{\prime}v^{\prime}$, although it is underestimated for $v^+_{\rm rms}$ and $w^+_{\rm rms}$ compared to the reference DNS.
This is likely caused by the fact that $u^{\prime +}$ is more dominant than the other velocity components in turbulent channel flow; hence, the information of $u^{\prime +}$ may be extracted preferentially by the present autoencoder.
Because of the overcompression, the extra small model underestimates the all velocity components as shown in figure~\ref{fig_RMSCNN}.
It is also striking that the curves obtained by the AE generally show underestimation against the reference DNS curve.
This is because the present CNN-AE is trained through the $L_2$ minimization manner as stated in equation \ref{eq:CNNL2}, which leads to output near-zero value, i.e., the average value of fluctuation components.
In other words, to output near zero value can be regarded as the safety option for the ML model to reduce the loss function in this particular sense.
The results here suggest that the large and medium models can successfully map the high-dimensional flow field into the low-dimensional latent space, while the small model can capture the global trends of DNS data.

\begin{figure}
	\begin{tabular}{cc}
	    \begin{minipage}[t]{0.95\hsize}
		\includegraphics[width=0.90\textwidth]{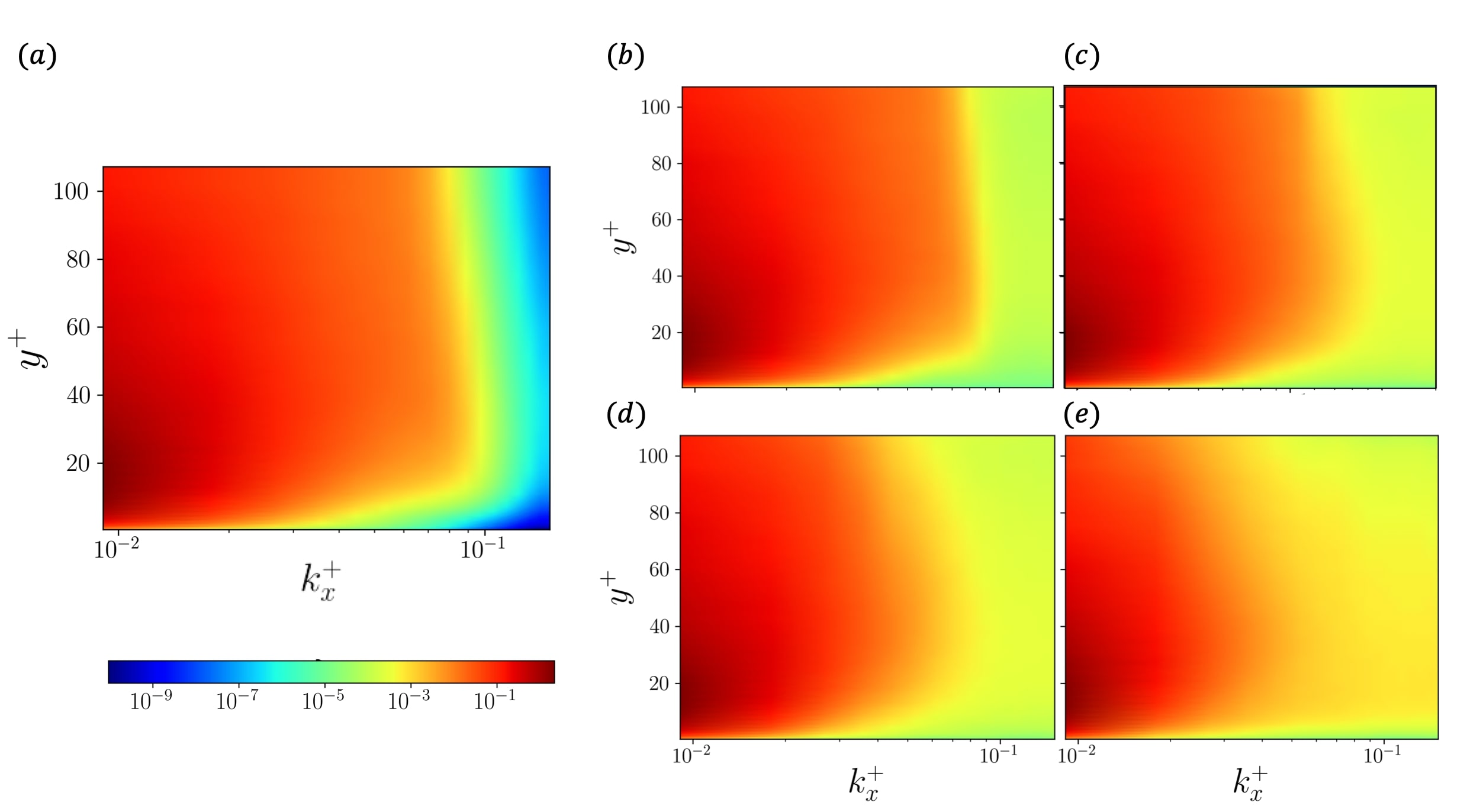}
		\caption{Pre-multiplied streamwise energy spectrum of flow field reconstructed by CNN-AE models. $(a)$ Reference DNS. $(b)$ Large. $(c)$ Medium. $(d)$ Small. $(e)$ Extra small.}
		\label{fig_CNNES_strpre}
	    \end{minipage}\\
	    
	    \begin{minipage}[t]{0.95\hsize}
		\includegraphics[width=0.90\textwidth]{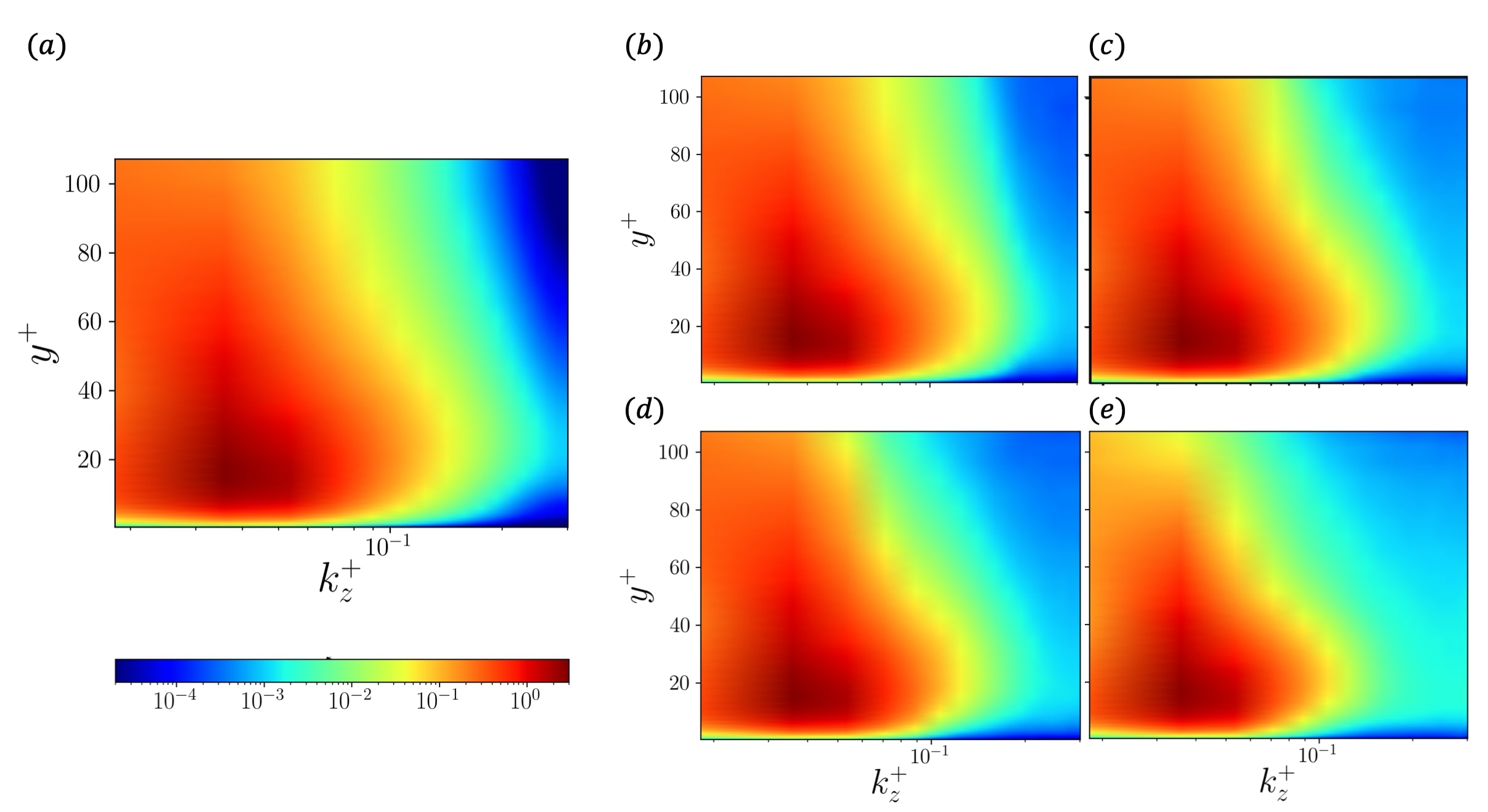}
		\caption{Pre-multiplied spanwise energy spectrum of flow field reconstructed by CNN-AE models. $(a)$ Reference DNS. $(b)$ Large. $(c)$ Medium. $(d)$ Small. $(e)$ Extra Small.
		}
		\label{fig_CNNES_spapre}
		\end{minipage}
	\end{tabular}
\end{figure}

We also assess the mapping ability of CNN-AE in the wavenumber space.
Let us respectively present in figures~\ref{fig_CNNES_strpre} and \ref{fig_CNNES_spapre} maps of the premultiplied streamwise and spanwise energy spectra of the streamwise velocity using the output fields of CNN-AE models.
Here, the one-dimensional
streamwise and spanwise spectra are defined as 
\begin{eqnarray}
    E_{uu}(k_x; y) &=& \overline{{\hat{u}}^\ast\hat{u}}^{z,t},\\
    E_{uu}(k_z; y) &=& \overline{{\hat{u}}^\ast\hat{u}}^{x,t},
\end{eqnarray}
where $(\cdot)^\ast$ represents the complex conjugate, and $\hat{(\cdot)}$ denotes a one-dimensional Fourier transform, respectively.
Note that the color contour in these figures is expressed using the log scale.
The energy spectrum for the large model shows good agreement with the reference DNS in the range of $k_x^+{,k_z^+} \lesssim 10^{-1}$, which suggests that {the model can cover low wavenumber components.}
The medium model is also able to reconstruct the low wavenumber components although the overestimated portion is larger than the large model as can been seen in figures~\ref{fig_CNNES_strpre}$(c)$ and \ref{fig_CNNES_spapre}$(c)$.
These trends are analogous to the large and medium models' ability to reconstruct small vortex structures as we have observed in figures~\ref{fig_QCNN}$(b)$ and $(c)$.
For the small and extra small models, the low wavenumber components can be captured well although the overestimation can be found at the high-wavenumber area due to the overcompression, which corresponds to the aforementioned trend that the flow field reconstructed by the extra small model has less vortical structures than by the small model as shown in figures~\ref{fig_QCNN}$(d),(e)$.
These observations indicate that the present CNN-AE tends to map energetically dominant components, i.e., low-wavenumber components, into the latent space preferentially as the size of latent vector is getting smaller.

\begin{figure}
	\begin{center}
		\includegraphics[width=0.45\textwidth]{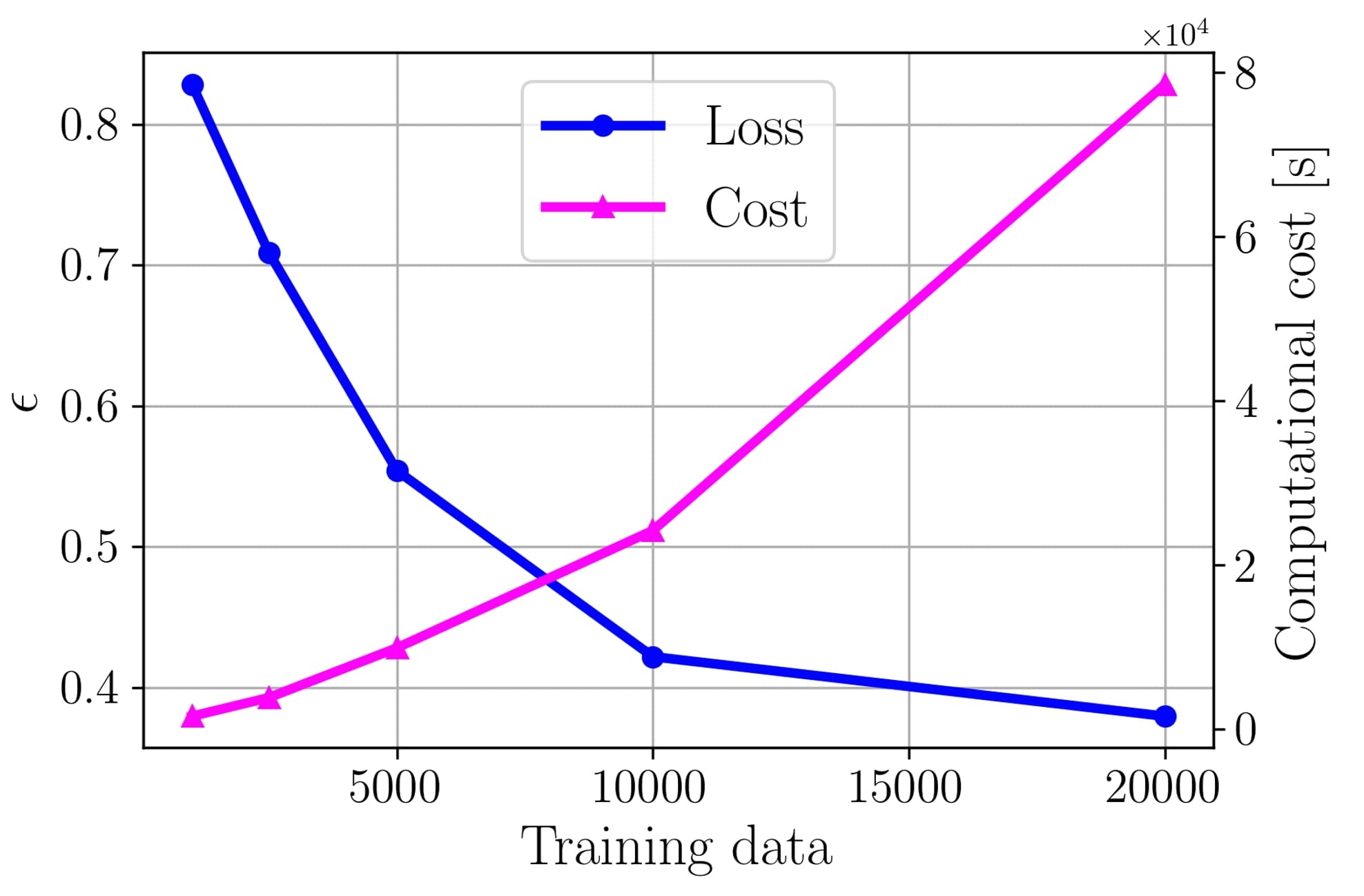}
		\caption{Dependence of the ability of CNN-AE on the amount of training snapshots.}
		\label{fig_costCNN}
	\end{center}
\end{figure}

Let us here investigate the dependence of mapping ability by the CNN-AE on the amount of training snapshots.
As an example, we consider the small model whose size of the latent vector is $(4,4,4,3)$.
The $L_2$ error norm between the input and output of the models {$\epsilon=||{\bm q}-{\cal F}_C({\bm q})||_2/||{\bm q}||_2$} and the computational cost on GPU (NVIDIA Tesla V100) for training are summarized in figure~\ref{fig_costCNN}.
Note that the plots on the computation cost do not need to increase linearly due to the use of early stopping criteria~\cite{prechelt1998}, which can prevent an overfitting.
The error and computational cost are in a trade-off relationship with each other; thus, the number of training snapshots should be decided taking in consideration of error requirements and computational resources.
The increase in the amount of training data considerably decreases the $L_2$ error norm up to 10\,000 training snapshots; however, the error reduction does not grow well from 10\,000 to 20\,000 training snapshots compared to the low number of snapshots.
Therefore, we have used 10\,000 snapshots as the training data for all cases.

\begin{figure}
	\begin{center}
		\includegraphics[width=0.98\textwidth]{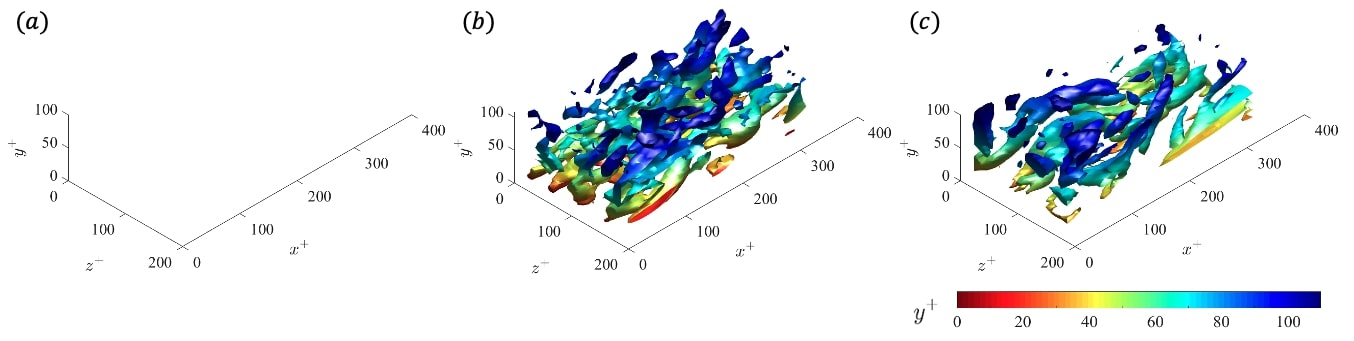}
		\caption{Isosurfaces of the second invariant of velocity gradient tensor ($Q^+=0.01$) at an arbitrary time 
instant{, computed using the test DNS data}.
		$(a)$ $t^{+}=2103$. $(b)$ $t^{+}=2739$. $(c)$ $t^{+}=4140$, where the time $t^{+}=0$ corresponds to the beginning of test data. 
		{Note that no vortical structure can be seen at $t^+=2103$.}
		}
		\label{fig_QsDNS}
	\end{center}
\end{figure}
\begin{figure}
	\begin{center}
		\includegraphics[width=0.50\textwidth]{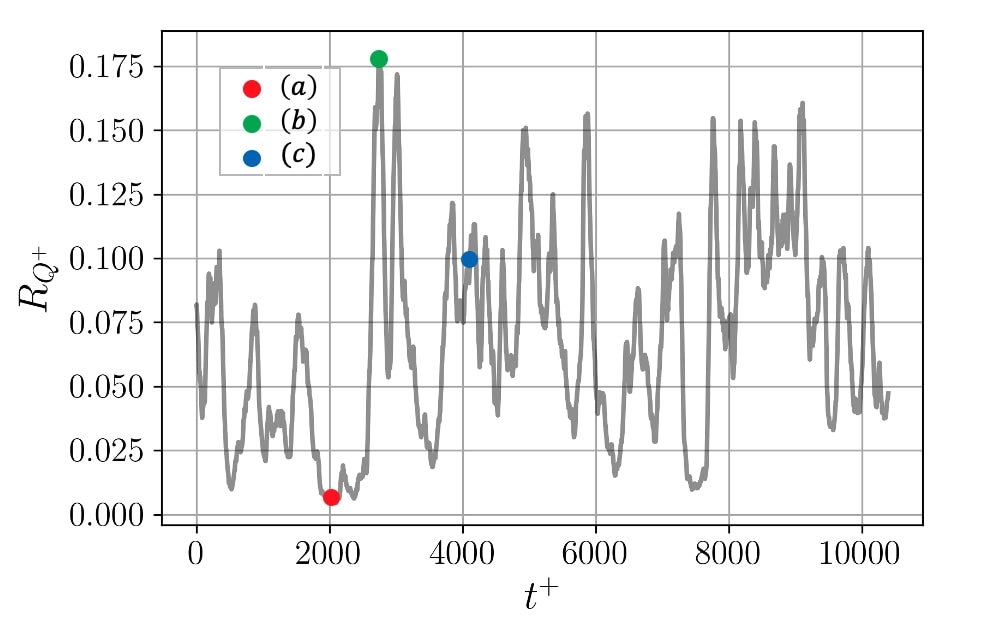}
		\caption{Temporal evolution of $R_{Q^+_{\rm vis}=0.01}$. The horizontal axis corresponds to the time 
		instant 
		in figure \ref{fig_QsDNS}.}
		\label{fig_RqtimeDNS}
	\end{center}
\end{figure}

\begin{figure}
	\begin{center}
		\includegraphics[width=0.70\textwidth]{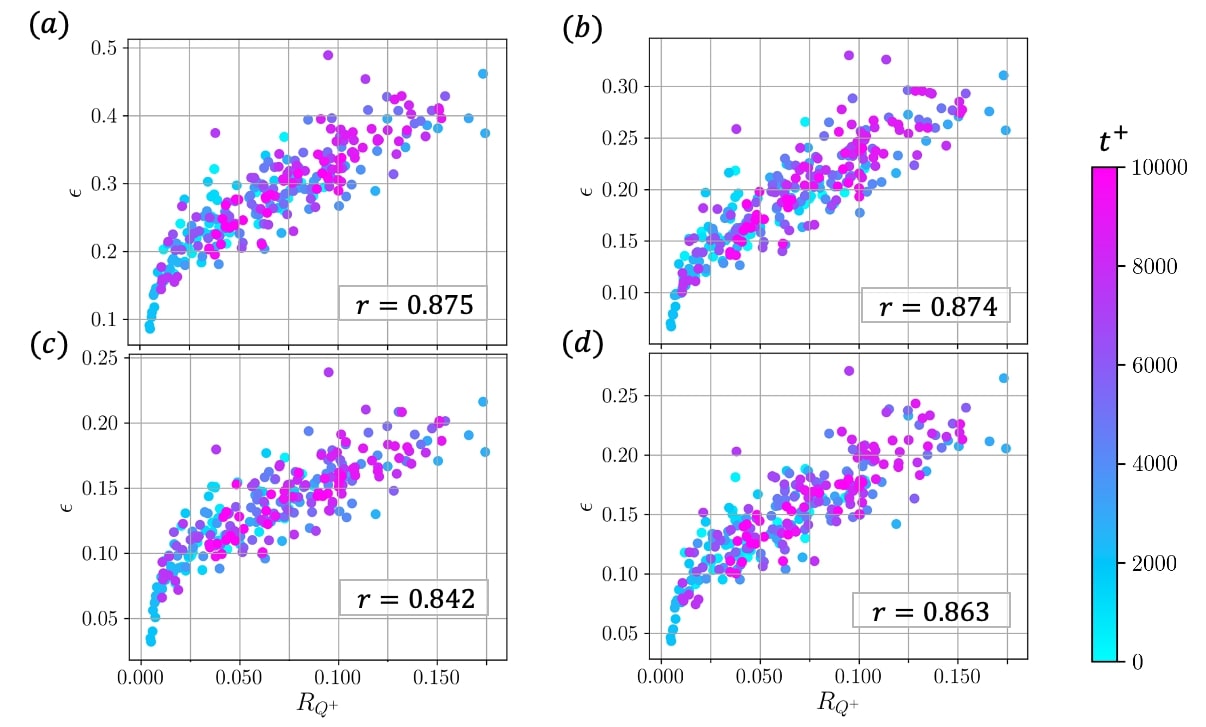}
		\caption{Correlation analysis between $R_{Q^+_{\rm vis}=0.01}$ and $L_2$ error of $(a)$ all velocity components, $(b)$ $u^{\prime +}$, $(c)$ $v^{\prime +}$, and $(d)$ $w^{\prime +}$. {The value of} $r$ represents the correlation coefficient.}
		\label{fig_RqECNN}
	\end{center}
\end{figure}
\begin{figure}
	\begin{center}
		\includegraphics[width=0.50\textwidth]{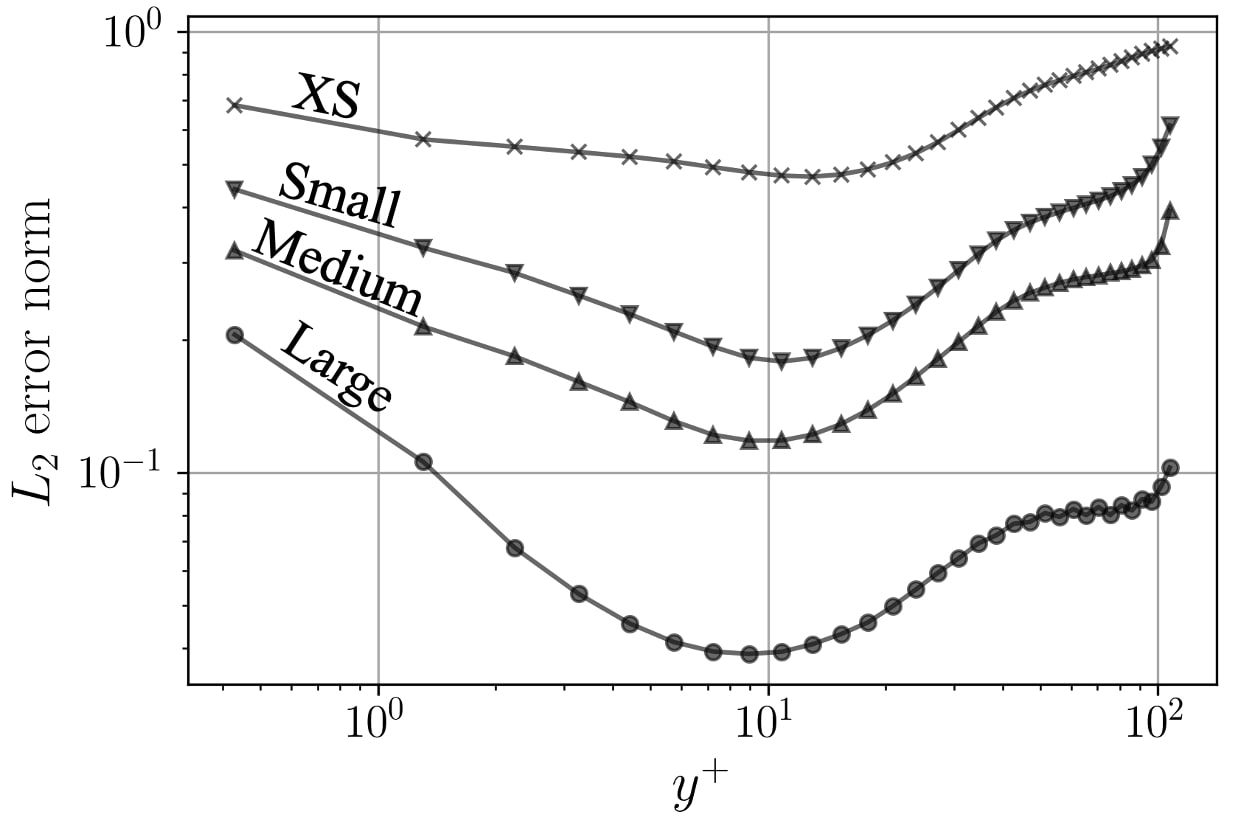}
		\caption{{The dependence of the $L_2$ error norm in the wall-normal direction.}}
		\label{fig_CNN_err_y}
	\end{center}
\end{figure}

In the discussion above, the time-ensemble averaged statistics were considered to analyze the mapping ability of the CNN-AE.
We can also suspect that the performance of the CNN-AE may also depend on each instantaneous state in turbulent channel flow since the amount of vortices further varies in time, as shown in figure \ref{fig_QsDNS}.
Here, let us investigate the relationship between the existence of strong vortex cores and the ability of CNN-AE.
To examine this point quantitatively, we define the ratio of number of grid points visualized by the second invariant of velocity gradient tensor to the total number of grid points:
\begin{equation}
    R_{Q^+_{\rm vis}} = \frac{n_{Q^+_{\rm vis}}}{n_{Q^+_{\rm vis}} + n_{Q^+_{\rm inv}}},
\end{equation}
where $n_{Q^+_{\rm vis}}$ and $n_{Q^+_{\rm inv}}$ are defined as
\begin{eqnarray}
    n_{Q^+_{\rm vis}} = n\bigl(\{Q^+ \mid  Q^+ < Q^+_{th}\}\bigr),\\
    n_{Q^+_{\rm inv}} = n\bigl(\{Q^+ \mid  Q^+ > Q^+_{th}\}\bigr),
\end{eqnarray}
using the threshold $Q^+_{th}$.
Note that $n(\{A\})$ represents the number of elements which is a member of the set $\{A\}$.
The temporal evolution of $R_{Q^+_{\rm vis}}$ over the test data is shown in figure~\ref{fig_RqtimeDNS}.
As can be seen, the ratio of visualized vortices constantly changes in time, similarly to the observation in figure~\ref{fig_QsDNS}.

Based on the finding above, we check the dependence of the CNN-AE performance on the instantaneous amount of vortex considering the correlation between the $L_2$ error norm and $R_{Q^+_{\rm vis}}$ as summarized in figure~\ref{fig_RqECNN}.
We here use $Q^+_{\rm vis}=0.01$ as the criteria.
As shown, there is a strong positive correlation between $R_{Q^+_{\rm vis}}$ and the errors computed by both all velocity components and each attribute.
It implies that the CNN-AE is better at reconstructing the flow field when there are no strong vortex cores in the flow.

{The dependence of the $L_2$ error norm on the $y$ position is also investigated, as shown in figure~\ref{fig_CNN_err_y}.
What is notable here is the high error level near the wall for all models.
This is likely due to the low-probability events in the near-wall region, i.e., the high flatness factor~\cite{KMM1987,FFT2020b}.
Since the present CNN-AEs are trained in the $L_2$ minimization manner as stated above, it is a tougher task to estimate near-wall region than the other region with less low-probability events.
Note that the observation here is consistent with that in figure~\ref{fig_RqECNN}, which indicates that the CNN-AE is good at estimating the flow field with no strong vortex cores.
}

\subsection{LSTM based temporal prediction on low-dimensional space}

\begin{figure}
	\begin{center}
		\includegraphics[width=0.80\textwidth]{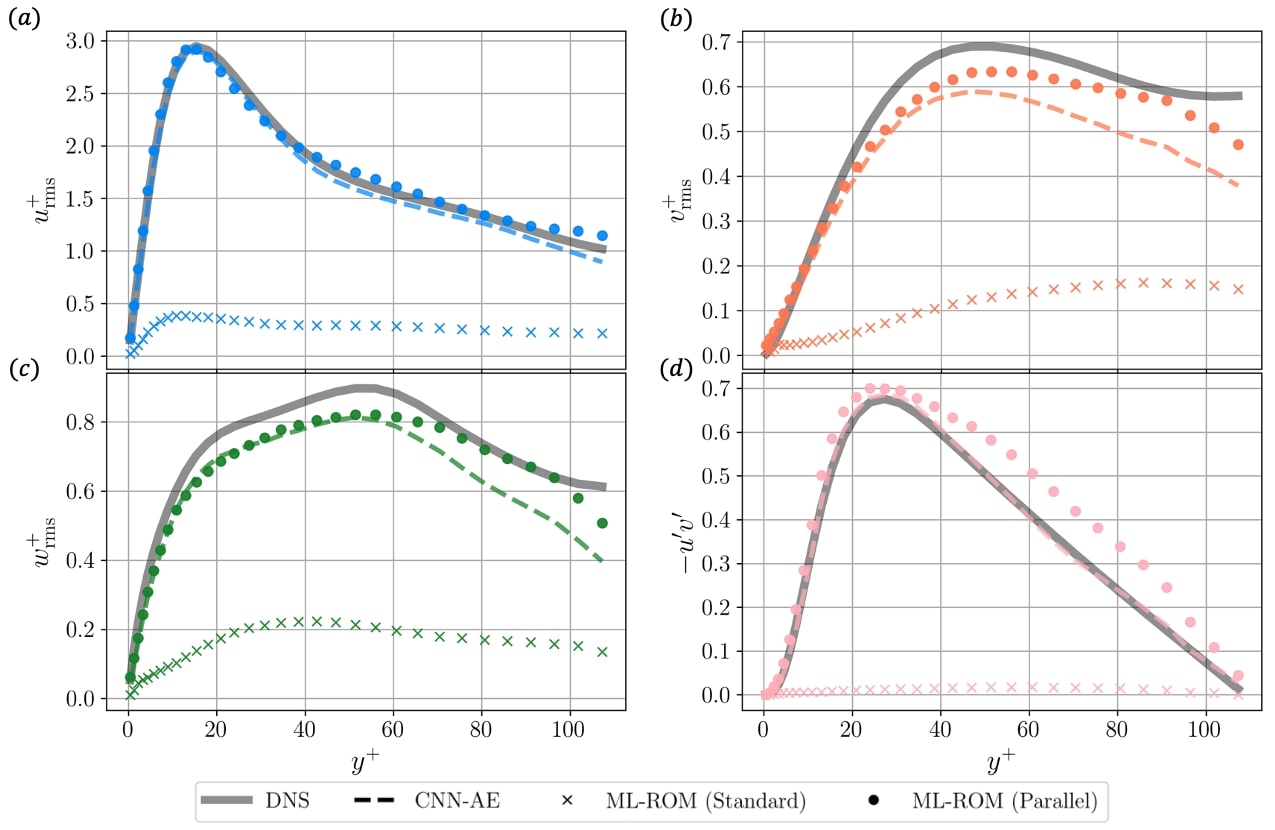}
		\caption{{Statistics of reconstructed flow field by the ML-ROM. $(a)$ $u^{+}_{\rm rms}$. $(b)$ $v^{+}_{\rm rms}$. $(c)$ $w^{+}_{\rm rms}$. $(d)$ $-\overline{u^{\prime +}v^{\prime +}}$. `Standard' and `Parallel' respectively indicate the two types of LSTM model used for the construction of ML-ROM.
		}}
		\label{fig_RMSLSTM}
	\end{center}
\end{figure}

As introduced in section~\ref{sec:CNN-LSTM}, the latent vectors obtained by the CNN-AE are then fed into the LSTM model to predict the time evolution of dynamics of flow fields.
Since the temporal evolution of high-dimensional fields can be obtained from only following that in a low-dimensional latent space, it can be regarded as reduced order modeling.
In this section, we use the medium CNN-AE model, which can capture the trends of DNS as already discussed in the previous section, so as to prepare the training data for the LSTM.
Analogous to the CNN-AE training, we use 10\,000 realizations obtained by reference DNS data and the CNN-AE for the LSTM training.
Again, latent vectors at five previous time steps are used as the input to predict that at the next time step such that ${\bm \eta}^{n\Delta t}={\cal F}_L(\{{\bm \eta}^{(n-1)\Delta t},{\bm \eta}^{(n-2)\Delta t},{\bm \eta}^{(n-3)\Delta t},{\bm \eta}^{(n-4)\Delta t},{\bm \eta}^{(n-5)\Delta t}\})$.
The Adam algorithm~\cite{kingma2014} is applied as the optimizer for weight updating, and early stopping~\cite{prechelt1998} is applied to train the models and avoid overfitting.

The generated output by the ML-ROM is first assessed with some statistics as shown in figure~\ref{fig_RMSLSTM}, which summarizes the RMS of all velocity attributes and Reynolds shear stress $-\overline{u^\prime v^\prime}$.
{The statistics are accumulated in a time period of about $10\,000$ wall unit time, which corresponds to $2\,600$ internal iterations in the LSTM model.}
{For comparison, we consider two types of LSTM
model --- 1. a standard LSTM with 768 units and 2. a parallel LSTM introduced in figure~\ref{fig_LSTM}$(b)$.}
The curve obtained by the medium CNN-AE only is also shown as the dashed line for comparison.
{Except for ML-ROM with the standard LSTM, the} statistics are in reasonable agreement with the reference DNS curve, although the result of ML-ROM is overestimated near center of the channel.
This is likely due to the influence of the recursive input in the sense that the output at the current time step is utilized as the input at the next time step.

\begin{figure}
	\begin{center}
		\includegraphics[width=0.85\textwidth]{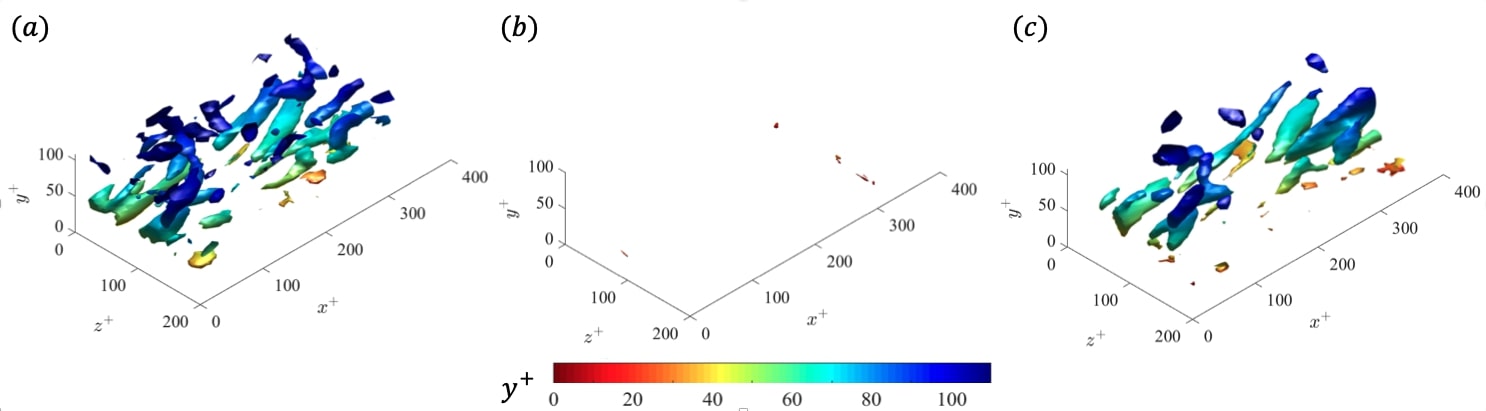}
		\caption{Instantaneous isosurfaces of the second invariant of velocity gradient tensor ($Q^+=0.01$). $(a)$ DNS. $(b)$ ML-ROM { with the standard LSTM. $(c)$ ML-ROM with the parallel LSTM.}}
		\label{fig_QLSTM}
	\end{center}
\end{figure}

The instantaneous flow field reconstructed by the ML-ROM is visualized using the second invariant of velocity gradient tensor ($Q^+=0.01$), as shown in figure~\ref{fig_QLSTM}.
The flow field generated by the ML-ROM {{with the parallel LSTM}} contains various scales of vortex structures, 
{which can also be found in the DNS flow field.
Based on the results in figures~\ref{fig_RMSLSTM} and \ref{fig_QLSTM}, we hereafter consider only the use of the parallel LSTM for the construction of the present ML-ROM.
}

The results in figures~\ref{fig_RMSLSTM} and \ref{fig_QLSTM} enable us to suspect that the flow field obtained by the present ML-ROM has the similar characteristics to the reference DNS although it does not match in the instantaneous sense.
To dig into this point, we consider the orbit in a phase space composed of the turbulent kinetic energy (TKE) $\int_V k dV$, the production of TKE $\int_V P dV$, and the dissipation of TKE $\int_V D dV$, which are integrated in the whole domain of the channel.
These values used for the construction of orbit are mathematically expressed as
\begin{eqnarray}
    \int_V P dV &=& \int_V -u^{\prime}v^{\prime} \frac{dU}{dy}dV,\\
    \int_V k dV &=& \int_V ({u^{\prime}}^2 + {v^{\prime}}^2 + {w^{\prime}}^2) dV,
\end{eqnarray}
\begin{eqnarray}
    \int_V D dV = \int_V \nu\biggl
(\frac{\partial u^{\prime}}{\partial x}\frac{\partial u^{\prime}}{\partial x}+\frac{\partial u^{\prime}}{\partial y}\frac{\partial u^{\prime}}{\partial y}+\frac{\partial u^{\prime}}{\partial z}\frac{\partial u^{\prime}}{\partial z}+\frac{\partial v^{\prime}}{\partial x}\frac{\partial v^{\prime}}{\partial x}+\frac{\partial v^{\prime}}{\partial y}\frac{\partial v^{\prime}}{\partial y}\nonumber\\
+\frac{\partial v^{\prime}}{\partial z}\frac{\partial v^{\prime}}{\partial z}
+\frac{\partial w^{\prime}}{\partial x}\frac{\partial w^{\prime}}{\partial x}+\frac{\partial w^{\prime}}{\partial y}\frac{\partial w^{\prime}}{\partial y}+\frac{\partial w^{\prime}}{\partial z}\frac{\partial w^{\prime}}{\partial z}\biggr) dV,
\end{eqnarray}
respectively, where $U$ is the mean streamwise velocity.
\begin{figure}
	\begin{center}
		\includegraphics[width=1.00\textwidth]{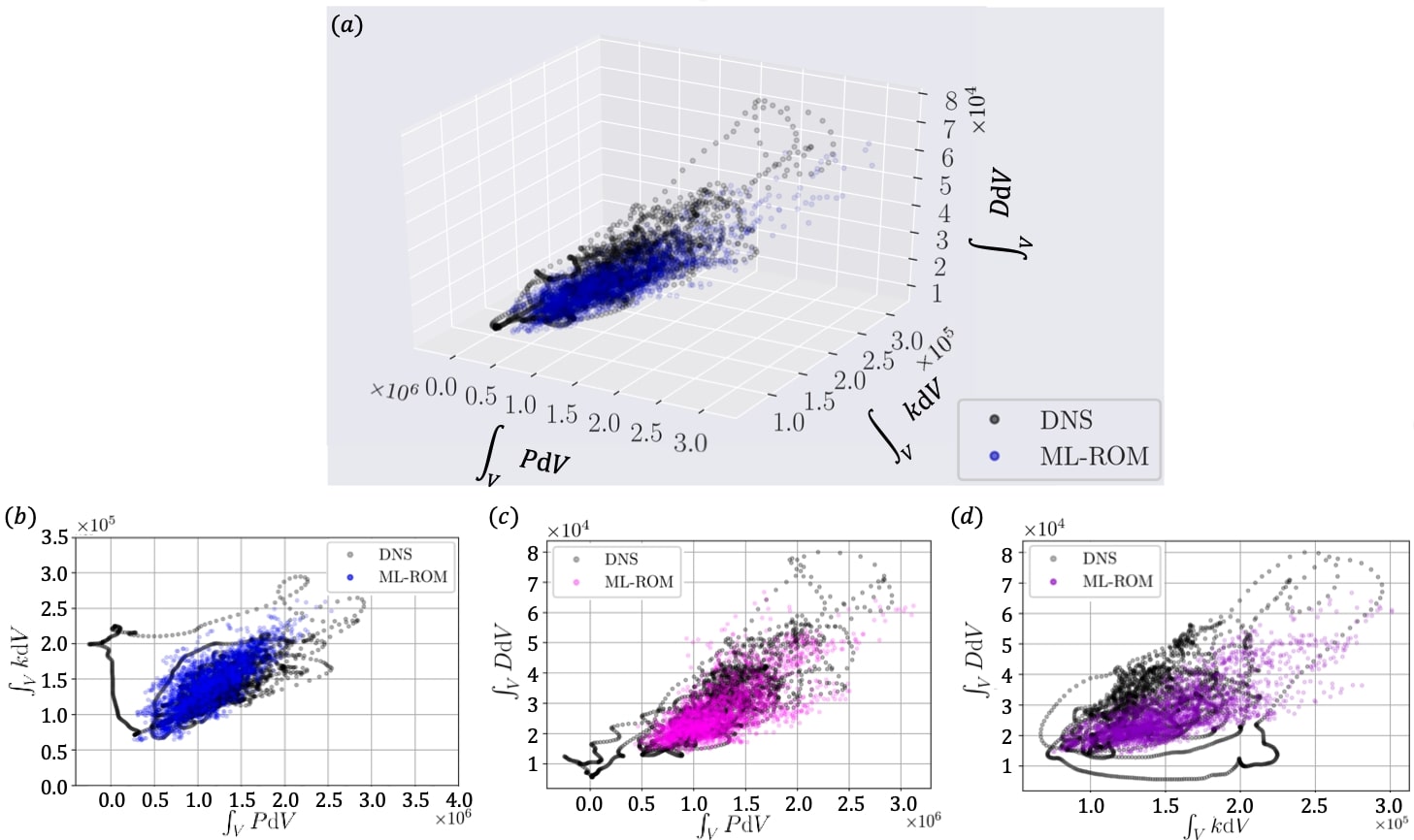}
		\caption{{Turbulence statistics-based orbits of reference DNS and ML-ROM. $(a)$ Three-dimensional orbit of TKE vs dissipation vs production. Two-dimensional orbits of $(b)$ TKE and production of TKE, $(c)$ TKE and dissipation, and $(d)$ production and dissipation.}}
		\label{fig_orbit}
	\end{center}
\end{figure}
The relationship among $\int_V P dV$, $\int_V k dV$ and $\int_V D dV$ is shown in figure~\ref{fig_orbit}.
In the aspect of time sequence, the orbit of ML-ROM is distinct against the reference DNS; however, the area of orbit is overlapping, which implies that the attractor may exist in the similar position in the phase space.
It also suggests that the temporal behavior of ML-ROM based flow fields is akin to that of the DNS.
Note that this observation is similar to Srinivasan {\it et al}.~\cite{SGASV2019} who used the LSTM to predict coefficients of low-dimensional turbulent shear flow model.

\begin{table}[]
\caption{ML-ROM models for investigation of dependence on time step.}
\begin{center}
\begin{tabular}{cccc}
\hline
\hline
Case        & \multicolumn{1}{l}{$\Delta t^+$} & \multicolumn{1}{l}{$\Delta t^+/\Delta t^+_{\rm DNS}$} & $R_{uu}^+(t^+)/R_{uu}^+(0)$\\ \hline
M-Short     & 3.85 & 100 & { 0.870}                           \\ 
M-Medium    & 7.70 & 200 & { 0.846}                             \\ 
M-Wide      & 15.4 & 400 & { 0.713}                             \\ 
M-Superwide (M-SW) & 30.8 & 800 & { 0.497}                            \\ 
\hline\hline
\end{tabular}
\end{center}
\label{table2}
\end{table}

\begin{figure}
	\begin{center}
		\includegraphics[width=0.45\textwidth]{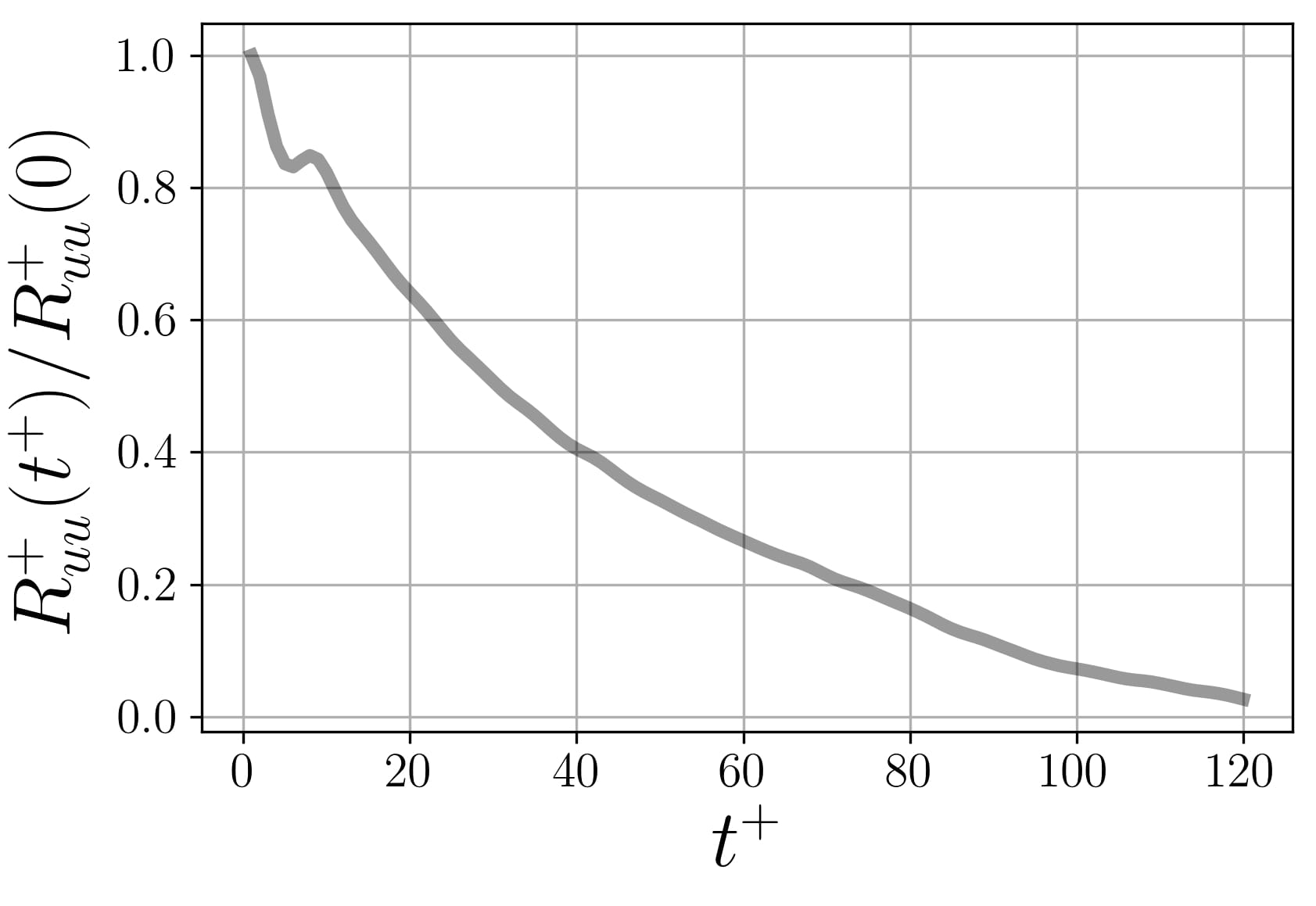}
		\caption{
		Temporal two-point correlation coefficient near the wall ($y^+=13.0$).
		}
		\label{fig_timecorr}
	\end{center}
\end{figure}

In addition to the accuracy of the present ML-ROM, its other considerable advantage against DNS or large-eddy simulation is that the ML-ROM may be able to predict the longer time interval than conventional simulations.
Here, let us investigate the dependence of prediction performance by the ML-ROM on the time interval for the LSTM.
Four different ML-ROM models, which contain the common medium CNN-AE with four different LSTMs, are considered as summarized in Table~\ref{table2}.
The Case M-Short is the same model as that for the discussion above, whose LSTM is trained with the time interval of $\Delta t^+=3.85$.
Note in passing that the time interval for the M-Short (baseline model) is already 100 folds compared to the DNS as shown in Table~\ref{table2}.
{We also present the temporal two-point correlation coefficient $R^+_{uu}(t^+)/R^+_{uu}(t^+=0)$ near the wall $(y^+=13.0)$ in figure~\ref{fig_timecorr}.
The coefficients for each LSTM model are also shown in Table~\ref{table2}.
It is known that the performance of machine learning model for turbulence in the temporal direction highly depends on the temporal correlation~\cite{FFT2020b}.
Although this is just one of the examples, the combination with correlation assessment enables us to choose the ML-ROM parameters appropriately in constructing surrogate models since the present ML-ROM is based on a supervised learning.
}

\begin{figure}
	\begin{center}
		\includegraphics[width=0.85\textwidth]{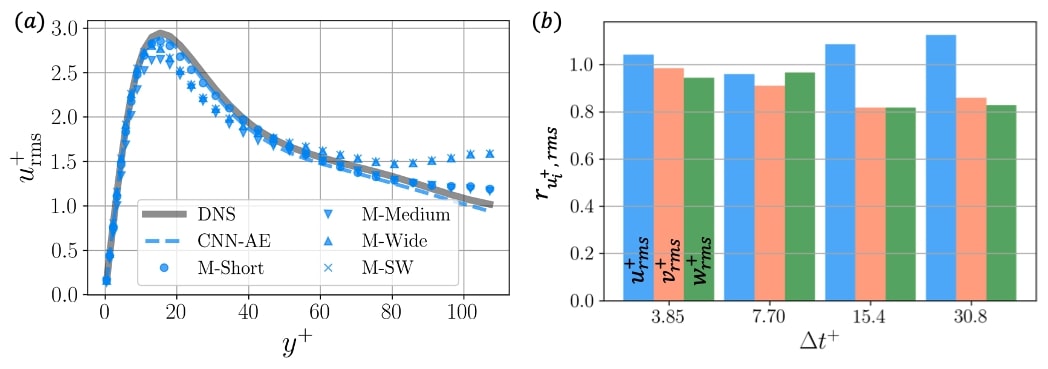}
		\caption{{$(a)$ Root-mean-squared (RMS) value of streamwise velocity component $u^{+}_{\rm rms}$. The curve obtained by the medium CNN-AE model is shown as the dashed line for comparison.} $(b)$ Time-ensemble peak error $r_{u^+_{i,{\rm rms}}}$ of ML-ROM models at the covered time steps.}
		\label{fig_tsRMSMLROM}
	\end{center}
\end{figure}

The dependence on the time interval for the LSTM is first compared using the RMS of streamwise velocity components, as presented in figure~\ref{fig_tsRMSMLROM}$(a)$.
The curves of M-Short and M-Medium are in reasonable agreement with 
that of the reference DNS.
In contrast, the effect of long time interval can be observed with the M-Wide and M-Superwide (M-SW) cases which shows the overestimation at $y^+\geq 20$, although the rough trends can be captured.
Note that the similar trend has been observed in both the $v$ and $w$ components although not shown here.
To check the influence on the time interval explicitly, we also use the error in the peak value of RMS $u^+_{i,{\rm rms}}$ of each velocity component in figure~\ref{fig_tsRMSMLROM}$(b)$,
\begin{equation}
    r_{u^+_{i,{\rm rms}}} = \frac{{\rm Max}[\overline{u^+_{i,{\rm rms},{\rm ML}}}^t]}{{\rm Max}[\overline{u^+_{i,{\rm rms},{\rm DNS}}}^t]}. \label{eq:rurms}
\end{equation}
Here, the time-ensemble average is taken over 10\,000 viscous time to obtain each term in equation~\ref{eq:rurms}.
This measure can report that the peak of velocity RMS by the ML-ROM is close to that by the DNS when $r_{u^+_{i,{\rm rms}}}\approx 1$.
The general trend observed here is that the error increases with the time step as we can expect.

\begin{figure}
	\begin{center}
		\includegraphics[width=0.95\textwidth]{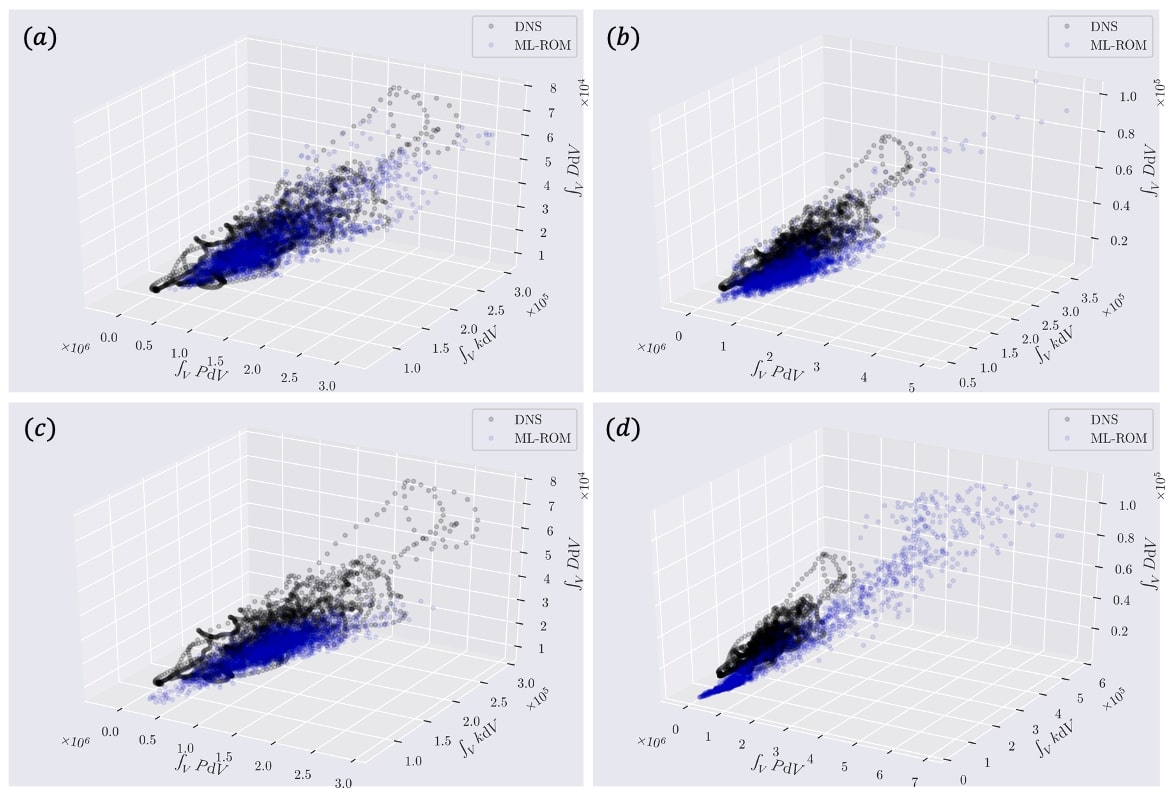}
		\caption{Three-dimensional orbits of the reference DNS and $(a)$ M-Short, $(b)$ M-Medium, $(c)$ M-Wide, and $(d)$ M-Superwide.}
		\label{fig_tsatMLROM}
	\end{center}
\end{figure}

The aforementioned trends can also be found with the orbit-based analysis.
As shown in figures~\ref{fig_tsatMLROM} $(a)$ and $(b)$, the distribution of M-Short and M-Medium shows the overlap with the reference DNS, which suggests that the ML-ROM is able to represent chaotic temporal behavior like the reference simulation.
On the other hand, because of the wide time interval, the dissipation of M-Superwide presented in figure~\ref{fig_tsatMLROM}$(d)$ further overestimates, which results in the different orbit behavior against the reference.
It implies that a proper choice of time step should be required to accurately reproduce the chaotic behavior, which is akin to the DNS, by the present ML-ROM.

\section{Concluding remarks}
\label{sec:con}

We examined the applicability of machine learning based reduced order model (ML-ROM) proposed by Hasegawa {\it et al.}~\cite{HFMF2020a} to three-dimensional complex flows.
A three-dimensional convolutional neural network autoencoder (CNN-AE) was employed to map a high-dimensional flow field into a low-dimensional latent space, and a long short-term memory (LSTM) was utilized to predict temporal evolution of the latent vector.
As an example, a turbulent channel flow at $Re_{\tau}=110$ in a minimal domain was considered.
The flow fields reconstructed by the ML-ROM were in good statistical agreement with the reference DNS data in time-ensemble sense.
The fidelity of the ML-ROM based turbulence could also be found according to the orbit-based analysis.

Moreover, additional case studies were conducted to investigate the dependence on parameters used for the present ML-ROM.
The investigation of the dependence on size of latent space suggested that {approximately 1\,500 modes} are required for the CNN-AE to reconstruct the flow field.
We also found that too wide time step size of LSTM makes it difficult for the ML-ROM to reconstruct flow field like DNS because of a low temporal correlation, suggesting that a careful choice should be 
made.

Although the present ML-ROM was able to represent flow fields like the reference DNS data, the instantaneous fields obtained by the ML-ROM do not completely match with the reference.
This is caused due to the recursive input of latent vector in the LSTM part, which is analogous to the observation in Srinivasan {\it et al}~\cite{SGASV2019}.
Necessity of a large number of latent modes is also one of the limitations in the present ML-ROM formation toward the applications to more complex flows, e.g., higher Reynolds number turbulence.
To tackle this issue, Fukami {\it et al.}~\cite{FNF2020} have recently introduced the use of hierarchical autoencoder which can achieve more efficient low-dimensionalization than the conventional AEs and POD.
This idea may be combined to the present ML-ROM scheme though their demonstration was performed with a two-dimensional sectional velocity of turbulent flow.
Incorporating customized loss functions so that represented flows satisfy physics laws {and capture small-scale structures} can also be considered~\cite{LY2019,raissi2020hidden,MFRFT2020,buzzicotti2020reconstruction}.
{In addition, the combination with other unsupervised schemes, e.g., generative adversarial network (GAN), is considerable for the applications of machine learning methods to flows at practically
high Reynolds numbers~\cite{iwamoto2002reynolds,yamamoto2018numerical}.
For instance, Kim and Lee~\cite{kim2020deep} proposed a GAN-based temporal integrator for turbulence.
Besides, Kim {\it et al.}~\cite{kim2020unsupervised} have recently reported the possibility of the use of GAN for the super-resolution analysis of high-Reynolds number flows.
Their framework can work well even for the test Reynolds numbers which are set to be much higher than that for training.
Moreover, the adaptability for unstructured mesh data is also favorable direction since the present CNN is based on a filter operation on a regular grid, which can be addressed by capitalizing on the state-of-the art CNNs~\cite{FukamiVoronoi,kashefi2020point,ogoke2020graph,tencer2020enabling,morimoto2101convolutional}.}
For the perspective on stability of neural network-based time integration, the concept of multiscale hierarchical time stepping may also be helpful~\cite{LKB2020}. 
Although the aforementioned extensions are just examples, we believe that the present report can serve as a significant step toward the establishment of machine learning based reduced order models of chaotic turbulence phenomena.

\section*{Acknowledgements}
This work was supported through JSPS KAKENHI Grant Number 18H03758.  
The authors acknowledge Mr. Masaki Morimoto (Keio Univ.) for fruitful discussions.
Kai Fukami also thanks Dr. Romit Maulik (Argonne National Lab.) for insightful comments with regard to autoencoder-based analysis of fluid flows.

\section*{Data Availability}
The data that support the findings of this study are available from the corresponding author upon reasonable request.

\section*{References}


\bibliography{NFHNF2020}

\begin{thebibliography}{65}%
\makeatletter
\providecommand \@ifxundefined [1]{%
 \@ifx{#1\undefined}
}%
\providecommand \@ifnum [1]{%
 \ifnum #1\expandafter \@firstoftwo
 \else \expandafter \@secondoftwo
 \fi
}%
\providecommand \@ifx [1]{%
 \ifx #1\expandafter \@firstoftwo
 \else \expandafter \@secondoftwo
 \fi
}%
\providecommand \natexlab [1]{#1}%
\providecommand \enquote  [1]{``#1''}%
\providecommand \bibnamefont  [1]{#1}%
\providecommand \bibfnamefont [1]{#1}%
\providecommand \citenamefont [1]{#1}%
\providecommand \href@noop [0]{\@secondoftwo}%
\providecommand \href [0]{\begingroup \@sanitize@url \@href}%
\providecommand \@href[1]{\@@startlink{#1}\@@href}%
\providecommand \@@href[1]{\endgroup#1\@@endlink}%
\providecommand \@sanitize@url [0]{\catcode `\\12\catcode `\$12\catcode
  `\&12\catcode `\#12\catcode `\^12\catcode `\_12\catcode `\%12\relax}%
\providecommand \@@startlink[1]{}%
\providecommand \@@endlink[0]{}%
\providecommand \url  [0]{\begingroup\@sanitize@url \@url }%
\providecommand \@url [1]{\endgroup\@href {#1}{\urlprefix }}%
\providecommand \urlprefix  [0]{URL }%
\providecommand \Eprint [0]{\href }%
\providecommand \doibase [0]{https://doi.org/}%
\providecommand \selectlanguage [0]{\@gobble}%
\providecommand \bibinfo  [0]{\@secondoftwo}%
\providecommand \bibfield  [0]{\@secondoftwo}%
\providecommand \translation [1]{[#1]}%
\providecommand \BibitemOpen [0]{}%
\providecommand \bibitemStop [0]{}%
\providecommand \bibitemNoStop [0]{.\EOS\space}%
\providecommand \EOS [0]{\spacefactor3000\relax}%
\providecommand \BibitemShut  [1]{\csname bibitem#1\endcsname}%
\let\auto@bib@innerbib\@empty
\bibitem [{\citenamefont {Lumley}(1967)}]{Lumely1967}%
  \BibitemOpen
  \bibfield  {author} {\bibinfo {author} {\bibfnamefont {J.~L.}\ \bibnamefont
  {Lumley}},\ }\bibfield  {title} {\enquote {\bibinfo {title} {The structure of
  inhomogeneous turbulent flows},}\ }in\ \href@noop {} {\emph {\bibinfo
  {booktitle} {Atmospheric turbulence and radio wave propagation}}},\ \bibinfo
  {editor} {edited by\ \bibinfo {editor} {\bibfnamefont {A.~M.}\ \bibnamefont
  {Yaglom}}\ and\ \bibinfo {editor} {\bibfnamefont {V.~I.}\ \bibnamefont
  {Tatarski}}}\ (\bibinfo  {publisher} {Nauka},\ \bibinfo {year}
  {1967})\BibitemShut {NoStop}%
\bibitem [{\citenamefont {Taira}\ \emph {et~al.}(2017)\citenamefont {Taira},
  \citenamefont {Brunton}, \citenamefont {Dawson}, \citenamefont {Rowley},
  \citenamefont {Colonius}, \citenamefont {McKeon}, \citenamefont {Schmidt},
  \citenamefont {Gordeyev}, \citenamefont {Theofilis},\ and\ \citenamefont
  {Ukeiley}}]{TBDRCMSGTU2017}%
  \BibitemOpen
  \bibfield  {author} {\bibinfo {author} {\bibfnamefont {K.}~\bibnamefont
  {Taira}}, \bibinfo {author} {\bibfnamefont {S.~L.}\ \bibnamefont {Brunton}},
  \bibinfo {author} {\bibfnamefont {S.~T.~M.}\ \bibnamefont {Dawson}}, \bibinfo
  {author} {\bibfnamefont {C.~W.}\ \bibnamefont {Rowley}}, \bibinfo {author}
  {\bibfnamefont {T.}~\bibnamefont {Colonius}}, \bibinfo {author}
  {\bibfnamefont {B.~J.}\ \bibnamefont {McKeon}}, \bibinfo {author}
  {\bibfnamefont {O.~T.}\ \bibnamefont {Schmidt}}, \bibinfo {author}
  {\bibfnamefont {S.}~\bibnamefont {Gordeyev}}, \bibinfo {author}
  {\bibfnamefont {V.}~\bibnamefont {Theofilis}},\ and\ \bibinfo {author}
  {\bibfnamefont {L.~S.}\ \bibnamefont {Ukeiley}},\ }\bibfield  {title}
  {\enquote {\bibinfo {title} {Modal analysis of fluid flows: An overview},}\
  }\href@noop {} {\bibfield  {journal} {\bibinfo  {journal} {AIAA J.}\ }\textbf
  {\bibinfo {volume} {55}},\ \bibinfo {pages} {4013--4041} (\bibinfo {year}
  {2017})}\BibitemShut {NoStop}%
\bibitem [{\citenamefont {Taira}\ \emph {et~al.}(2020)\citenamefont {Taira},
  \citenamefont {Hemati}, \citenamefont {Brunton}, \citenamefont {Sun},
  \citenamefont {Duraisamy}, \citenamefont {Bagheri}, \citenamefont {Dawson},\
  and\ \citenamefont {Yeh}}]{THBSDBDY2019}%
  \BibitemOpen
  \bibfield  {author} {\bibinfo {author} {\bibfnamefont {K.}~\bibnamefont
  {Taira}}, \bibinfo {author} {\bibfnamefont {M.~S.}\ \bibnamefont {Hemati}},
  \bibinfo {author} {\bibfnamefont {S.~L.}\ \bibnamefont {Brunton}}, \bibinfo
  {author} {\bibfnamefont {Y.}~\bibnamefont {Sun}}, \bibinfo {author}
  {\bibfnamefont {K.}~\bibnamefont {Duraisamy}}, \bibinfo {author}
  {\bibfnamefont {S.}~\bibnamefont {Bagheri}}, \bibinfo {author} {\bibfnamefont
  {S.}~\bibnamefont {Dawson}},\ and\ \bibinfo {author} {\bibfnamefont {C.-A.}\
  \bibnamefont {Yeh}},\ }\bibfield  {title} {\enquote {\bibinfo {title} {Modal
  analysis of fluid flows: Applications and outlook},}\ }\href@noop {}
  {\bibfield  {journal} {\bibinfo  {journal} {AIAA J.}\ }\textbf {\bibinfo
  {volume} {58}},\ \bibinfo {pages} {998--1022} (\bibinfo {year}
  {2020})}\BibitemShut {NoStop}%
\bibitem [{\citenamefont {Noack}\ \emph {et~al.}(2003)\citenamefont {Noack},
  \citenamefont {Afanasiev}, \citenamefont {Morzynski}, \citenamefont
  {Tadmor},\ and\ \citenamefont {Thiele}}]{NAMTT2003}%
  \BibitemOpen
  \bibfield  {author} {\bibinfo {author} {\bibfnamefont {B.~R.}\ \bibnamefont
  {Noack}}, \bibinfo {author} {\bibfnamefont {K.}~\bibnamefont {Afanasiev}},
  \bibinfo {author} {\bibfnamefont {M.}~\bibnamefont {Morzynski}}, \bibinfo
  {author} {\bibfnamefont {G.}~\bibnamefont {Tadmor}},\ and\ \bibinfo {author}
  {\bibfnamefont {F.}~\bibnamefont {Thiele}},\ }\bibfield  {title} {\enquote
  {\bibinfo {title} {A hierarchy of low-dimensional models for the transient
  and post-transient cylinder wake},}\ }\href@noop {} {\bibfield  {journal}
  {\bibinfo  {journal} {J. Fluid Mech.}\ }\textbf {\bibinfo {volume} {497}},\
  \bibinfo {pages} {335–363} (\bibinfo {year} {2003})}\BibitemShut {NoStop}%
\bibitem [{\citenamefont {Noack}, \citenamefont {Papas},\ and\ \citenamefont
  {Monkewitz}(2005)}]{NPM2005}%
  \BibitemOpen
  \bibfield  {author} {\bibinfo {author} {\bibfnamefont {B.~R.}\ \bibnamefont
  {Noack}}, \bibinfo {author} {\bibfnamefont {P.}~\bibnamefont {Papas}},\ and\
  \bibinfo {author} {\bibfnamefont {P.~A.}\ \bibnamefont {Monkewitz}},\
  }\bibfield  {title} {\enquote {\bibinfo {title} {The need for a pressure-term
  representation in empirical galerkin models of incompressible shear flows},}\
  }\href@noop {} {\bibfield  {journal} {\bibinfo  {journal} {J. Fluid Mech.}\
  }\textbf {\bibinfo {volume} {523}},\ \bibinfo {pages} {339–365} (\bibinfo
  {year} {2005})}\BibitemShut {NoStop}%
\bibitem [{\citenamefont {Alfonsi}\ and\ \citenamefont
  {Primavera}(2007)}]{alfonsi2006}%
  \BibitemOpen
  \bibfield  {author} {\bibinfo {author} {\bibfnamefont {G.}~\bibnamefont
  {Alfonsi}}\ and\ \bibinfo {author} {\bibfnamefont {L.}~\bibnamefont
  {Primavera}},\ }\bibfield  {title} {\enquote {\bibinfo {title} {The structure
  of turbulent boundary layers in the wall region of plane channel flow},}\
  }\href@noop {} {\bibfield  {journal} {\bibinfo  {journal} {Proc. R. Soc. A}\
  }\textbf {\bibinfo {volume} {463}},\ \bibinfo {pages} {593--612} (\bibinfo
  {year} {2007})}\BibitemShut {NoStop}%
\bibitem [{\citenamefont {Muralidhar}\ \emph {et~al.}(2019)\citenamefont
  {Muralidhar}, \citenamefont {Podvin}, \citenamefont {Mathelin},\ and\
  \citenamefont {Fraigneau}}]{MPMF2019}%
  \BibitemOpen
  \bibfield  {author} {\bibinfo {author} {\bibfnamefont {S.~D.}\ \bibnamefont
  {Muralidhar}}, \bibinfo {author} {\bibfnamefont {B.}~\bibnamefont {Podvin}},
  \bibinfo {author} {\bibfnamefont {L.}~\bibnamefont {Mathelin}},\ and\
  \bibinfo {author} {\bibfnamefont {Y.}~\bibnamefont {Fraigneau}},\ }\bibfield
  {title} {\enquote {\bibinfo {title} {Spatio-temporal proper orthogonal
  decomposition of turbulent channel flow},}\ }\href@noop {} {\bibfield
  {journal} {\bibinfo  {journal} {J. Fluid Mech.}\ }\textbf {\bibinfo {volume}
  {864}},\ \bibinfo {pages} {614--639} (\bibinfo {year} {2019})}\BibitemShut
  {NoStop}%
\bibitem [{\citenamefont {Ilak}\ and\ \citenamefont
  {Rowley}(2008)}]{ilak2008modeling}%
  \BibitemOpen
  \bibfield  {author} {\bibinfo {author} {\bibfnamefont {M.}~\bibnamefont
  {Ilak}}\ and\ \bibinfo {author} {\bibfnamefont {C.~W.}\ \bibnamefont
  {Rowley}},\ }\bibfield  {title} {\enquote {\bibinfo {title} {Modeling of
  transitional channel flow using balanced proper orthogonal decomposition},}\
  }\href@noop {} {\bibfield  {journal} {\bibinfo  {journal} {Phys. Fluids}\
  }\textbf {\bibinfo {volume} {20}},\ \bibinfo {pages} {034103} (\bibinfo
  {year} {2008})}\BibitemShut {NoStop}%
\bibitem [{\citenamefont {Yu}, \citenamefont {Yan},\ and\ \citenamefont
  {Guo}(2019)}]{yu2019non}%
  \BibitemOpen
  \bibfield  {author} {\bibinfo {author} {\bibfnamefont {J.}~\bibnamefont
  {Yu}}, \bibinfo {author} {\bibfnamefont {C.}~\bibnamefont {Yan}},\ and\
  \bibinfo {author} {\bibfnamefont {M.}~\bibnamefont {Guo}},\ }\bibfield
  {title} {\enquote {\bibinfo {title} {Non-intrusive reduced-order modeling for
  fluid problems: A brief review},}\ }\href@noop {} {\bibfield  {journal}
  {\bibinfo  {journal} {J. Aerosp. Eng.}\ }\textbf {\bibinfo {volume} {233}},\
  \bibinfo {pages} {5896--5912} (\bibinfo {year} {2019})}\BibitemShut {NoStop}%
\bibitem [{\citenamefont {San}\ and\ \citenamefont {Maulik}(2018)}]{SM2018}%
  \BibitemOpen
  \bibfield  {author} {\bibinfo {author} {\bibfnamefont {O.}~\bibnamefont
  {San}}\ and\ \bibinfo {author} {\bibfnamefont {R.}~\bibnamefont {Maulik}},\
  }\bibfield  {title} {\enquote {\bibinfo {title} {Extreme learning machine for
  reduced order modeling of turbulent geophysical flows},}\ }\href@noop {}
  {\bibfield  {journal} {\bibinfo  {journal} {Phys. Rev. E}\ }\textbf {\bibinfo
  {volume} {97}},\ \bibinfo {pages} {04322} (\bibinfo {year}
  {2018})}\BibitemShut {NoStop}%
\bibitem [{\citenamefont {Pawar}\ \emph {et~al.}(2019)\citenamefont {Pawar},
  \citenamefont {Rahman}, \citenamefont {Vaddireddy}, \citenamefont {San},
  \citenamefont {Rasheed},\ and\ \citenamefont {Vedula}}]{pawar2019deep}%
  \BibitemOpen
  \bibfield  {author} {\bibinfo {author} {\bibfnamefont {S.}~\bibnamefont
  {Pawar}}, \bibinfo {author} {\bibfnamefont {S.~M.}\ \bibnamefont {Rahman}},
  \bibinfo {author} {\bibfnamefont {H.}~\bibnamefont {Vaddireddy}}, \bibinfo
  {author} {\bibfnamefont {O.}~\bibnamefont {San}}, \bibinfo {author}
  {\bibfnamefont {A.}~\bibnamefont {Rasheed}},\ and\ \bibinfo {author}
  {\bibfnamefont {P.}~\bibnamefont {Vedula}},\ }\bibfield  {title} {\enquote
  {\bibinfo {title} {A deep learning enabler for nonintrusive reduced order
  modeling of fluid flows},}\ }\href@noop {} {\bibfield  {journal} {\bibinfo
  {journal} {Phys. Fluids}\ }\textbf {\bibinfo {volume} {31}},\ \bibinfo
  {pages} {085101} (\bibinfo {year} {2019})}\BibitemShut {NoStop}%
\bibitem [{\citenamefont {Renganathan}, \citenamefont {Maulik},\ and\
  \citenamefont {Rao}(2020)}]{renganathan2020machine}%
  \BibitemOpen
  \bibfield  {author} {\bibinfo {author} {\bibfnamefont {S.~A.}\ \bibnamefont
  {Renganathan}}, \bibinfo {author} {\bibfnamefont {R.}~\bibnamefont
  {Maulik}},\ and\ \bibinfo {author} {\bibfnamefont {V.}~\bibnamefont {Rao}},\
  }\bibfield  {title} {\enquote {\bibinfo {title} {Machine learning for
  nonintrusive model order reduction of the parametric inviscid transonic flow
  past an airfoil},}\ }\href@noop {} {\bibfield  {journal} {\bibinfo  {journal}
  {Phys. Fluids}\ }\textbf {\bibinfo {volume} {32}},\ \bibinfo {pages} {047110}
  (\bibinfo {year} {2020})}\BibitemShut {NoStop}%
\bibitem [{\citenamefont {Maulik}\ \emph
  {et~al.}(2020{\natexlab{a}})\citenamefont {Maulik}, \citenamefont {Fukami},
  \citenamefont {Ramachandra}, \citenamefont {Fukagata},\ and\ \citenamefont
  {Taira}}]{MFRFT2020}%
  \BibitemOpen
  \bibfield  {author} {\bibinfo {author} {\bibfnamefont {R.}~\bibnamefont
  {Maulik}}, \bibinfo {author} {\bibfnamefont {K.}~\bibnamefont {Fukami}},
  \bibinfo {author} {\bibfnamefont {N.}~\bibnamefont {Ramachandra}}, \bibinfo
  {author} {\bibfnamefont {K.}~\bibnamefont {Fukagata}},\ and\ \bibinfo
  {author} {\bibfnamefont {K.}~\bibnamefont {Taira}},\ }\bibfield  {title}
  {\enquote {\bibinfo {title} {Probabilistic neural networks for fluid flow
  surrogate modeling and data recovery},}\ }\href@noop {} {\bibfield  {journal}
  {\bibinfo  {journal} {Phys. Rev. Fluids}\ }\textbf {\bibinfo {volume} {5}},\
  \bibinfo {pages} {104401} (\bibinfo {year} {2020}{\natexlab{a}})}\BibitemShut
  {NoStop}%
\bibitem [{\citenamefont {Srinivasan}\ \emph {et~al.}(2019)\citenamefont
  {Srinivasan}, \citenamefont {Guastoni}, \citenamefont {Azizpour},
  \citenamefont {Schlatter},\ and\ \citenamefont {Vinuesa}}]{SGASV2019}%
  \BibitemOpen
  \bibfield  {author} {\bibinfo {author} {\bibfnamefont {P.~A.}\ \bibnamefont
  {Srinivasan}}, \bibinfo {author} {\bibfnamefont {L.}~\bibnamefont
  {Guastoni}}, \bibinfo {author} {\bibfnamefont {H.}~\bibnamefont {Azizpour}},
  \bibinfo {author} {\bibfnamefont {P.}~\bibnamefont {Schlatter}},\ and\
  \bibinfo {author} {\bibfnamefont {R.}~\bibnamefont {Vinuesa}},\ }\bibfield
  {title} {\enquote {\bibinfo {title} {Predictions of turbulent shear flows
  using deep neural networks},}\ }\href@noop {} {\bibfield  {journal} {\bibinfo
   {journal} {Phys. Rev. Fluids}\ }\textbf {\bibinfo {volume} {4}},\ \bibinfo
  {pages} {054603} (\bibinfo {year} {2019})}\BibitemShut {NoStop}%
\bibitem [{\citenamefont {Hochreiter}\ and\ \citenamefont
  {Schmidhuber}(1997)}]{HS1997}%
  \BibitemOpen
  \bibfield  {author} {\bibinfo {author} {\bibfnamefont {S.}~\bibnamefont
  {Hochreiter}}\ and\ \bibinfo {author} {\bibfnamefont {J.}~\bibnamefont
  {Schmidhuber}},\ }\bibfield  {title} {\enquote {\bibinfo {title} {Long
  short-term memory},}\ }\href@noop {} {\bibfield  {journal} {\bibinfo
  {journal} {Neural Comput.}\ }\textbf {\bibinfo {volume} {9}},\ \bibinfo
  {pages} {1735--1780} (\bibinfo {year} {1997})}\BibitemShut {NoStop}%
\bibitem [{\citenamefont {Pawar}\ \emph {et~al.}(2020)\citenamefont {Pawar},
  \citenamefont {Ahmed}, \citenamefont {San},\ and\ \citenamefont
  {Rasheed}}]{pawar2020data}%
  \BibitemOpen
  \bibfield  {author} {\bibinfo {author} {\bibfnamefont {S.}~\bibnamefont
  {Pawar}}, \bibinfo {author} {\bibfnamefont {S.~E.}\ \bibnamefont {Ahmed}},
  \bibinfo {author} {\bibfnamefont {O.}~\bibnamefont {San}},\ and\ \bibinfo
  {author} {\bibfnamefont {A.}~\bibnamefont {Rasheed}},\ }\bibfield  {title}
  {\enquote {\bibinfo {title} {Data-driven recovery of hidden physics in
  reduced order modeling of fluid flows},}\ }\href@noop {} {\bibfield
  {journal} {\bibinfo  {journal} {Phys. Fluids}\ }\textbf {\bibinfo {volume}
  {32}},\ \bibinfo {pages} {036602} (\bibinfo {year} {2020})}\BibitemShut
  {NoStop}%
\bibitem [{\citenamefont {Hinton}\ and\ \citenamefont
  {Salakhutdinov}(2006)}]{HS2006}%
  \BibitemOpen
  \bibfield  {author} {\bibinfo {author} {\bibfnamefont {G.~E.}\ \bibnamefont
  {Hinton}}\ and\ \bibinfo {author} {\bibfnamefont {R.~R.}\ \bibnamefont
  {Salakhutdinov}},\ }\bibfield  {title} {\enquote {\bibinfo {title} {Reducing
  the dimensionality of data with neural networks},}\ }\href@noop {} {\bibfield
   {journal} {\bibinfo  {journal} {Science}\ }\textbf {\bibinfo {volume}
  {313}},\ \bibinfo {pages} {504--507} (\bibinfo {year} {2006})}\BibitemShut
  {NoStop}%
\bibitem [{\citenamefont {Brunton}, \citenamefont {Noack},\ and\ \citenamefont
  {Koumoutsakos}(2020)}]{BNK2020}%
  \BibitemOpen
  \bibfield  {author} {\bibinfo {author} {\bibfnamefont {S.~L.}\ \bibnamefont
  {Brunton}}, \bibinfo {author} {\bibfnamefont {B.~R.}\ \bibnamefont {Noack}},\
  and\ \bibinfo {author} {\bibfnamefont {P.}~\bibnamefont {Koumoutsakos}},\
  }\bibfield  {title} {\enquote {\bibinfo {title} {Machine learning for fluid
  mechanincs},}\ }\href@noop {} {\bibfield  {journal} {\bibinfo  {journal}
  {Annu. Rev. Fluid Mech.}\ }\textbf {\bibinfo {volume} {52}},\ \bibinfo
  {pages} {477--508} (\bibinfo {year} {2020})}\BibitemShut {NoStop}%
\bibitem [{\citenamefont {Brenner}, \citenamefont {Eldredge},\ and\
  \citenamefont {Freund}(2019)}]{BEF2019}%
  \BibitemOpen
  \bibfield  {author} {\bibinfo {author} {\bibfnamefont {M.~P.}\ \bibnamefont
  {Brenner}}, \bibinfo {author} {\bibfnamefont {J.~D.}\ \bibnamefont
  {Eldredge}},\ and\ \bibinfo {author} {\bibfnamefont {J.~B.}\ \bibnamefont
  {Freund}},\ }\bibfield  {title} {\enquote {\bibinfo {title} {Perspective on
  machine learning for advancing fluid mechanics},}\ }\href@noop {} {\bibfield
  {journal} {\bibinfo  {journal} {Phys. Rev. Fluids}\ }\textbf {\bibinfo
  {volume} {4}},\ \bibinfo {pages} {(100501)} (\bibinfo {year}
  {2019})}\BibitemShut {NoStop}%
\bibitem [{\citenamefont {Brunton}, \citenamefont {Hemanti},\ and\
  \citenamefont {Taira}(2020)}]{BHT2020}%
  \BibitemOpen
  \bibfield  {author} {\bibinfo {author} {\bibfnamefont {S.~L.}\ \bibnamefont
  {Brunton}}, \bibinfo {author} {\bibfnamefont {M.~S.}\ \bibnamefont
  {Hemanti}},\ and\ \bibinfo {author} {\bibfnamefont {K.}~\bibnamefont
  {Taira}},\ }\bibfield  {title} {\enquote {\bibinfo {title} {Special issue on
  machine learning and data-driven methods in fluid dynamics},}\ }\href@noop {}
  {\bibfield  {journal} {\bibinfo  {journal} {Theor. Comput. Fluid Dyn.}\
  }\textbf {\bibinfo {volume} {34}},\ \bibinfo {pages} {333--337} (\bibinfo
  {year} {2020})}\BibitemShut {NoStop}%
\bibitem [{\citenamefont {Fukami}, \citenamefont {Murata},\ and\ \citenamefont
  {Fukagata}(2020)}]{FMF2020}%
  \BibitemOpen
  \bibfield  {author} {\bibinfo {author} {\bibfnamefont {K.}~\bibnamefont
  {Fukami}}, \bibinfo {author} {\bibfnamefont {T.}~\bibnamefont {Murata}},\
  and\ \bibinfo {author} {\bibfnamefont {K.}~\bibnamefont {Fukagata}},\
  }\bibfield  {title} {\enquote {\bibinfo {title} {Sparse identification of
  nonlinear dynamics with low-dimensionalized flow representations},}\
  }\href@noop {} {\bibfield  {journal} {\bibinfo  {journal}
  {\rm{arXiv:2010.12177}\hspace{-0.3em}}\ } (\bibinfo {year}
  {2020})}\BibitemShut {NoStop}%
\bibitem [{\citenamefont {Milano}\ and\ \citenamefont
  {Koumoutsakos}(2002)}]{Milano2002}%
  \BibitemOpen
  \bibfield  {author} {\bibinfo {author} {\bibfnamefont {M.}~\bibnamefont
  {Milano}}\ and\ \bibinfo {author} {\bibfnamefont {P.}~\bibnamefont
  {Koumoutsakos}},\ }\bibfield  {title} {\enquote {\bibinfo {title} {Neural
  network modeling for near wall turbulent flow},}\ }\href@noop {} {\bibfield
  {journal} {\bibinfo  {journal} {J. Comput. Phys.}\ }\textbf {\bibinfo
  {volume} {182}},\ \bibinfo {pages} {1--26} (\bibinfo {year}
  {2002})}\BibitemShut {NoStop}%
\bibitem [{\citenamefont {Fukami}, \citenamefont {Fukagata},\ and\
  \citenamefont {Taira}(2020)}]{FFT2019b}%
  \BibitemOpen
  \bibfield  {author} {\bibinfo {author} {\bibfnamefont {K.}~\bibnamefont
  {Fukami}}, \bibinfo {author} {\bibfnamefont {K.}~\bibnamefont {Fukagata}},\
  and\ \bibinfo {author} {\bibfnamefont {K.}~\bibnamefont {Taira}},\ }\bibfield
   {title} {\enquote {\bibinfo {title} {Assessment of supervised machine
  learning for fluid flows},}\ }\href@noop {} {\bibfield  {journal} {\bibinfo
  {journal} {Theor. Comp. Fluid Dyn.}\ }\textbf {\bibinfo {volume} {34}},\
  \bibinfo {pages} {497--519} (\bibinfo {year} {2020})}\BibitemShut {NoStop}%
\bibitem [{\citenamefont {Omata}\ and\ \citenamefont
  {Shirayama}(2019)}]{omata2019}%
  \BibitemOpen
  \bibfield  {author} {\bibinfo {author} {\bibfnamefont {N.}~\bibnamefont
  {Omata}}\ and\ \bibinfo {author} {\bibfnamefont {S.}~\bibnamefont
  {Shirayama}},\ }\bibfield  {title} {\enquote {\bibinfo {title} {A novel
  method of low-dimensional representation for temporal behavior of flow fields
  using deep autoencoder},}\ }\href {https://doi.org/10.1063/1.5067313}
  {\bibfield  {journal} {\bibinfo  {journal} {AIP Adv.}\ }\textbf {\bibinfo
  {volume} {9}},\ \bibinfo {pages} {015006} (\bibinfo {year}
  {2019})}\BibitemShut {NoStop}%
\bibitem [{\citenamefont {Murata}, \citenamefont {Fukami},\ and\ \citenamefont
  {Fukagata}(2020)}]{MFF2019}%
  \BibitemOpen
  \bibfield  {author} {\bibinfo {author} {\bibfnamefont {T.}~\bibnamefont
  {Murata}}, \bibinfo {author} {\bibfnamefont {K.}~\bibnamefont {Fukami}},\
  and\ \bibinfo {author} {\bibfnamefont {K.}~\bibnamefont {Fukagata}},\
  }\bibfield  {title} {\enquote {\bibinfo {title} {Nonlinear mode decomposition
  with convolutional neural networks for fluid dynamics},}\ }\href@noop {}
  {\bibfield  {journal} {\bibinfo  {journal} {J. Fluid Mech.}\ }\textbf
  {\bibinfo {volume} {882}},\ \bibinfo {pages} {A13} (\bibinfo {year}
  {2020})}\BibitemShut {NoStop}%
\bibitem [{\citenamefont {Fukami}, \citenamefont {Nakamura},\ and\
  \citenamefont {Fukagata}(2020)}]{FNF2020}%
  \BibitemOpen
  \bibfield  {author} {\bibinfo {author} {\bibfnamefont {K.}~\bibnamefont
  {Fukami}}, \bibinfo {author} {\bibfnamefont {T.}~\bibnamefont {Nakamura}},\
  and\ \bibinfo {author} {\bibfnamefont {K.}~\bibnamefont {Fukagata}},\
  }\bibfield  {title} {\enquote {\bibinfo {title} {Convolutional neural network
  based hierarchical autoencoder for nonlinear mode decomposition of fluid
  field data},}\ }\href@noop {} {\bibfield  {journal} {\bibinfo  {journal}
  {Phys. Fluids}\ }\textbf {\bibinfo {volume} {32}},\ \bibinfo {pages} {095110}
  (\bibinfo {year} {2020})}\BibitemShut {NoStop}%
\bibitem [{\citenamefont {Hasegawa}\ \emph
  {et~al.}(2020{\natexlab{a}})\citenamefont {Hasegawa}, \citenamefont {Fukami},
  \citenamefont {Murata},\ and\ \citenamefont {Fukagata}}]{HFMF2020a}%
  \BibitemOpen
  \bibfield  {author} {\bibinfo {author} {\bibfnamefont {K.}~\bibnamefont
  {Hasegawa}}, \bibinfo {author} {\bibfnamefont {K.}~\bibnamefont {Fukami}},
  \bibinfo {author} {\bibfnamefont {T.}~\bibnamefont {Murata}},\ and\ \bibinfo
  {author} {\bibfnamefont {K.}~\bibnamefont {Fukagata}},\ }\bibfield  {title}
  {\enquote {\bibinfo {title} {Machine-learning-based reduced-order modeling
  for unsteady flows around bluff bodies of various shapes},}\ }\href@noop {}
  {\bibfield  {journal} {\bibinfo  {journal} {Theor. Comp. Fluid Dyn.}\
  }\textbf {\bibinfo {volume} {34}},\ \bibinfo {pages} {367--388} (\bibinfo
  {year} {2020}{\natexlab{a}})}\BibitemShut {NoStop}%
\bibitem [{\citenamefont {Hasegawa}\ \emph {et~al.}(2019)\citenamefont
  {Hasegawa}, \citenamefont {Fukami}, \citenamefont {Murata},\ and\
  \citenamefont {Fukagata}}]{HFMF2019}%
  \BibitemOpen
  \bibfield  {author} {\bibinfo {author} {\bibfnamefont {K.}~\bibnamefont
  {Hasegawa}}, \bibinfo {author} {\bibfnamefont {K.}~\bibnamefont {Fukami}},
  \bibinfo {author} {\bibfnamefont {T.}~\bibnamefont {Murata}},\ and\ \bibinfo
  {author} {\bibfnamefont {K.}~\bibnamefont {Fukagata}},\ }\bibfield  {title}
  {\enquote {\bibinfo {title} {Data-driven reduced order modeling of flows
  around two-dimensional bluff bodies of various shapes},}\ }\href@noop {}
  {\bibfield  {journal} {\bibinfo  {journal} {ASME-JSME-KSME Joint Fluids
  Engineering Conference, San Francisco, USA}\ } (\bibinfo {year}
  {2019})}\BibitemShut {NoStop}%
\bibitem [{\citenamefont {Hasegawa}\ \emph
  {et~al.}(2020{\natexlab{b}})\citenamefont {Hasegawa}, \citenamefont {Fukami},
  \citenamefont {Murata},\ and\ \citenamefont {Fukagata}}]{HFMF2020b}%
  \BibitemOpen
  \bibfield  {author} {\bibinfo {author} {\bibfnamefont {K.}~\bibnamefont
  {Hasegawa}}, \bibinfo {author} {\bibfnamefont {K.}~\bibnamefont {Fukami}},
  \bibinfo {author} {\bibfnamefont {T.}~\bibnamefont {Murata}},\ and\ \bibinfo
  {author} {\bibfnamefont {K.}~\bibnamefont {Fukagata}},\ }\bibfield  {title}
  {\enquote {\bibinfo {title} {{CNN-LSTM} based reduced order modeling of
  two-dimensional unsteady flows around a circular cylinder at different
  reynolds numbers},}\ }\href {https://doi.org/10.1088/1873-7005/abb91d}
  {\bibfield  {journal} {\bibinfo  {journal} {Fluid. Dyn. Res.}\ } (\bibinfo
  {year} {2020}{\natexlab{b}}),\ 10.1088/1873-7005/abb91d}\BibitemShut
  {NoStop}%
\bibitem [{\citenamefont {Iwamoto}, \citenamefont {Suzuki},\ and\ \citenamefont
  {Kasagi}(2002)}]{iwamoto2002reynolds}%
  \BibitemOpen
  \bibfield  {author} {\bibinfo {author} {\bibfnamefont {K.}~\bibnamefont
  {Iwamoto}}, \bibinfo {author} {\bibfnamefont {Y.}~\bibnamefont {Suzuki}},\
  and\ \bibinfo {author} {\bibfnamefont {N.}~\bibnamefont {Kasagi}},\
  }\bibfield  {title} {\enquote {\bibinfo {title} {Reynolds number effect on
  wall turbulence: toward effective feedback control},}\ }\href@noop {}
  {\bibfield  {journal} {\bibinfo  {journal} {Int. J. Heat Fluid Flow}\
  }\textbf {\bibinfo {volume} {23}},\ \bibinfo {pages} {678--689} (\bibinfo
  {year} {2002})}\BibitemShut {NoStop}%
\bibitem [{\citenamefont {Jim{\'e}nez}\ and\ \citenamefont
  {Moin}(1991)}]{jimenez1991minimal}%
  \BibitemOpen
  \bibfield  {author} {\bibinfo {author} {\bibfnamefont {J.}~\bibnamefont
  {Jim{\'e}nez}}\ and\ \bibinfo {author} {\bibfnamefont {P.}~\bibnamefont
  {Moin}},\ }\bibfield  {title} {\enquote {\bibinfo {title} {The minimal flow
  unit in near-wall turbulence},}\ }\href@noop {} {\bibfield  {journal}
  {\bibinfo  {journal} {J. Fluid Mech.}\ }\textbf {\bibinfo {volume} {225}},\
  \bibinfo {pages} {213--240} (\bibinfo {year} {1991})}\BibitemShut {NoStop}%
\bibitem [{\citenamefont {Morinishi}\ \emph {et~al.}(1998)\citenamefont
  {Morinishi}, \citenamefont {Lund}, \citenamefont {Vasilyev},\ and\
  \citenamefont {Moin}}]{morinishi1998fully}%
  \BibitemOpen
  \bibfield  {author} {\bibinfo {author} {\bibfnamefont {Y.}~\bibnamefont
  {Morinishi}}, \bibinfo {author} {\bibfnamefont {T.~S.}\ \bibnamefont {Lund}},
  \bibinfo {author} {\bibfnamefont {O.~V.}\ \bibnamefont {Vasilyev}},\ and\
  \bibinfo {author} {\bibfnamefont {P.}~\bibnamefont {Moin}},\ }\bibfield
  {title} {\enquote {\bibinfo {title} {Fully conservative higher order finite
  difference schemes for incompressible flow},}\ }\href@noop {} {\bibfield
  {journal} {\bibinfo  {journal} {J. Comput. Phys.}\ }\textbf {\bibinfo
  {volume} {143}},\ \bibinfo {pages} {90--124} (\bibinfo {year}
  {1998})}\BibitemShut {NoStop}%
\bibitem [{\citenamefont {Spalart}, \citenamefont {Moser},\ and\ \citenamefont
  {Rogers}(1991)}]{spalart1991spectral}%
  \BibitemOpen
  \bibfield  {author} {\bibinfo {author} {\bibfnamefont {P.~R.}\ \bibnamefont
  {Spalart}}, \bibinfo {author} {\bibfnamefont {R.~D.}\ \bibnamefont {Moser}},\
  and\ \bibinfo {author} {\bibfnamefont {M.~M.}\ \bibnamefont {Rogers}},\
  }\bibfield  {title} {\enquote {\bibinfo {title} {Spectral methods for the
  navier-stokes equations with one infinite and two periodic directions},}\
  }\href@noop {} {\bibfield  {journal} {\bibinfo  {journal} {J. Comput. Phys.}\
  }\textbf {\bibinfo {volume} {96}},\ \bibinfo {pages} {297--324} (\bibinfo
  {year} {1991})}\BibitemShut {NoStop}%
\bibitem [{\citenamefont {Dukowicz}\ and\ \citenamefont
  {Dvinsky}(1992)}]{dukowicz1992approximate}%
  \BibitemOpen
  \bibfield  {author} {\bibinfo {author} {\bibfnamefont {J.~K.}\ \bibnamefont
  {Dukowicz}}\ and\ \bibinfo {author} {\bibfnamefont {A.~S.}\ \bibnamefont
  {Dvinsky}},\ }\bibfield  {title} {\enquote {\bibinfo {title} {Approximate
  factorization as a high order splitting for the implicit incompressible flow
  equations},}\ }\href@noop {} {\bibfield  {journal} {\bibinfo  {journal} {J.
  Comput. Phys.}\ }\textbf {\bibinfo {volume} {102}},\ \bibinfo {pages}
  {336--347} (\bibinfo {year} {1992})}\BibitemShut {NoStop}%
\bibitem [{\citenamefont {Shanker}, \citenamefont {Hu},\ and\ \citenamefont
  {Hung}(1996)}]{shanker1996effect}%
  \BibitemOpen
  \bibfield  {author} {\bibinfo {author} {\bibfnamefont {M.}~\bibnamefont
  {Shanker}}, \bibinfo {author} {\bibfnamefont {M.~Y.}\ \bibnamefont {Hu}},\
  and\ \bibinfo {author} {\bibfnamefont {M.~S.}\ \bibnamefont {Hung}},\
  }\bibfield  {title} {\enquote {\bibinfo {title} {Effect of data
  standardization on neural network training},}\ }\href@noop {} {\bibfield
  {journal} {\bibinfo  {journal} {Omega}\ }\textbf {\bibinfo {volume} {24}},\
  \bibinfo {pages} {385--397} (\bibinfo {year} {1996})}\BibitemShut {NoStop}%
\bibitem [{\citenamefont {LeCun}\ \emph {et~al.}(1998)\citenamefont {LeCun},
  \citenamefont {Bottou}, \citenamefont {Bengio},\ and\ \citenamefont
  {Haffner}}]{LBBH1998}%
  \BibitemOpen
  \bibfield  {author} {\bibinfo {author} {\bibfnamefont {Y.}~\bibnamefont
  {LeCun}}, \bibinfo {author} {\bibfnamefont {L.}~\bibnamefont {Bottou}},
  \bibinfo {author} {\bibfnamefont {Y.}~\bibnamefont {Bengio}},\ and\ \bibinfo
  {author} {\bibfnamefont {P.}~\bibnamefont {Haffner}},\ }\bibfield  {title}
  {\enquote {\bibinfo {title} {Gradient-based learning applied to document
  recognition},}\ }\href@noop {} {\bibfield  {journal} {\bibinfo  {journal}
  {Proc. IEEE}\ }\textbf {\bibinfo {volume} {86}},\ \bibinfo {pages}
  {2278--2324} (\bibinfo {year} {1998})}\BibitemShut {NoStop}%
\bibitem [{\citenamefont {Fukami}\ \emph {et~al.}(2019)\citenamefont {Fukami},
  \citenamefont {Nabae}, \citenamefont {Kawai},\ and\ \citenamefont
  {Fukagata}}]{FNKF2019}%
  \BibitemOpen
  \bibfield  {author} {\bibinfo {author} {\bibfnamefont {K.}~\bibnamefont
  {Fukami}}, \bibinfo {author} {\bibfnamefont {Y.}~\bibnamefont {Nabae}},
  \bibinfo {author} {\bibfnamefont {K.}~\bibnamefont {Kawai}},\ and\ \bibinfo
  {author} {\bibfnamefont {K.}~\bibnamefont {Fukagata}},\ }\bibfield  {title}
  {\enquote {\bibinfo {title} {Synthetic turbulent inflow generator using
  machine learning},}\ }\href@noop {} {\bibfield  {journal} {\bibinfo
  {journal} {Phys. Rev. Fluids}\ }\textbf {\bibinfo {volume} {4}},\ \bibinfo
  {pages} {064603} (\bibinfo {year} {2019})}\BibitemShut {NoStop}%
\bibitem [{\citenamefont {Fukami}, \citenamefont {Fukagata},\ and\
  \citenamefont {Taira}(2019{\natexlab{a}})}]{FFT2019a}%
  \BibitemOpen
  \bibfield  {author} {\bibinfo {author} {\bibfnamefont {K.}~\bibnamefont
  {Fukami}}, \bibinfo {author} {\bibfnamefont {K.}~\bibnamefont {Fukagata}},\
  and\ \bibinfo {author} {\bibfnamefont {K.}~\bibnamefont {Taira}},\ }\bibfield
   {title} {\enquote {\bibinfo {title} {Super-resolution reconstruction of
  turbulent flows with machine learning},}\ }\href
  {https://doi.org/10.1017/jfm.2019.238} {\bibfield  {journal} {\bibinfo
  {journal} {J. Fluid Mech.}\ }\textbf {\bibinfo {volume} {870}},\ \bibinfo
  {pages} {106--120} (\bibinfo {year} {2019}{\natexlab{a}})}\BibitemShut
  {NoStop}%
\bibitem [{\citenamefont {Fukami}, \citenamefont {Fukagata},\ and\
  \citenamefont {Taira}(2019{\natexlab{b}})}]{FFT2019tsfp}%
  \BibitemOpen
  \bibfield  {author} {\bibinfo {author} {\bibfnamefont {K.}~\bibnamefont
  {Fukami}}, \bibinfo {author} {\bibfnamefont {K.}~\bibnamefont {Fukagata}},\
  and\ \bibinfo {author} {\bibfnamefont {K.}~\bibnamefont {Taira}},\ }\bibfield
   {title} {\enquote {\bibinfo {title} {Super-resolution analysis with machine
  learning for low-resolution flow data},}\ }in\ \href@noop {} {\emph {\bibinfo
  {booktitle} {11th International Symposium on Turbulence and Shear Flow
  Phenomena (TSFP11), Southampton, UK}}},\ \bibinfo {series and number}
  {\bibinfo {number} {208}}\ (\bibinfo {year} {2019})\BibitemShut {NoStop}%
\bibitem [{\citenamefont {Champion}\ \emph {et~al.}(2019)\citenamefont
  {Champion}, \citenamefont {Lusch}, \citenamefont {Kutz},\ and\ \citenamefont
  {Brunton}}]{CLKB2019}%
  \BibitemOpen
  \bibfield  {author} {\bibinfo {author} {\bibfnamefont {K.}~\bibnamefont
  {Champion}}, \bibinfo {author} {\bibfnamefont {B.}~\bibnamefont {Lusch}},
  \bibinfo {author} {\bibfnamefont {J.~N.}\ \bibnamefont {Kutz}},\ and\
  \bibinfo {author} {\bibfnamefont {S.~L.}\ \bibnamefont {Brunton}},\
  }\bibfield  {title} {\enquote {\bibinfo {title} {Data-driven discovery of
  coordinates and governing equations},}\ }\href@noop {} {\bibfield  {journal}
  {\bibinfo  {journal} {Proc. Natl. Acad. Sci.}\ }\textbf {\bibinfo {volume}
  {45}},\ \bibinfo {pages} {116} (\bibinfo {year} {2019})}\BibitemShut
  {NoStop}%
\bibitem [{\citenamefont {Morimoto}, \citenamefont {Fukami},\ and\
  \citenamefont {Fukagata}(2020)}]{MFF2020}%
  \BibitemOpen
  \bibfield  {author} {\bibinfo {author} {\bibfnamefont {M.}~\bibnamefont
  {Morimoto}}, \bibinfo {author} {\bibfnamefont {K.}~\bibnamefont {Fukami}},\
  and\ \bibinfo {author} {\bibfnamefont {K.}~\bibnamefont {Fukagata}},\
  }\bibfield  {title} {\enquote {\bibinfo {title} {Experimental velocity data
  estimation for imperfect particle images using machine learning},}\
  }\href@noop {} {\bibfield  {journal} {\bibinfo  {journal}
  {\rm{arXiv:2005.00756}\hspace{-0.3em}}\ } (\bibinfo {year}
  {2020})}\BibitemShut {NoStop}%
\bibitem [{\citenamefont {Liu}\ \emph {et~al.}(2020)\citenamefont {Liu},
  \citenamefont {Tang}, \citenamefont {Huang},\ and\ \citenamefont
  {Lu}}]{LTHL2020}%
  \BibitemOpen
  \bibfield  {author} {\bibinfo {author} {\bibfnamefont {B.}~\bibnamefont
  {Liu}}, \bibinfo {author} {\bibfnamefont {J.}~\bibnamefont {Tang}}, \bibinfo
  {author} {\bibfnamefont {H.}~\bibnamefont {Huang}},\ and\ \bibinfo {author}
  {\bibfnamefont {X.-Y.}\ \bibnamefont {Lu}},\ }\bibfield  {title} {\enquote
  {\bibinfo {title} {Deep learning methods for super-resolution reconstruction
  of turbulent flows},}\ }\href@noop {} {\bibfield  {journal} {\bibinfo
  {journal} {Phys. Fluids}\ }\textbf {\bibinfo {volume} {32}},\ \bibinfo
  {pages} {025105} (\bibinfo {year} {2020})}\BibitemShut {NoStop}%
\bibitem [{\citenamefont {Kim}\ and\ \citenamefont
  {Lee}(2020{\natexlab{a}})}]{KL2020}%
  \BibitemOpen
  \bibfield  {author} {\bibinfo {author} {\bibfnamefont {J.}~\bibnamefont
  {Kim}}\ and\ \bibinfo {author} {\bibfnamefont {C.}~\bibnamefont {Lee}},\
  }\bibfield  {title} {\enquote {\bibinfo {title} {Prediction of turbulent heat
  transfer using convolutional neural networks},}\ }\href@noop {} {\bibfield
  {journal} {\bibinfo  {journal} {J. Fluid Mech.}\ }\textbf {\bibinfo {volume}
  {882}},\ \bibinfo {pages} {A18} (\bibinfo {year}
  {2020}{\natexlab{a}})}\BibitemShut {NoStop}%
\bibitem [{\citenamefont {Fukami}\ \emph {et~al.}(2020)\citenamefont {Fukami},
  \citenamefont {Hasegawa}, \citenamefont {Nakamura}, \citenamefont
  {Morimoto},\ and\ \citenamefont {Fukagata}}]{FHNMF2020}%
  \BibitemOpen
  \bibfield  {author} {\bibinfo {author} {\bibfnamefont {K.}~\bibnamefont
  {Fukami}}, \bibinfo {author} {\bibfnamefont {K.}~\bibnamefont {Hasegawa}},
  \bibinfo {author} {\bibfnamefont {T.}~\bibnamefont {Nakamura}}, \bibinfo
  {author} {\bibfnamefont {M.}~\bibnamefont {Morimoto}},\ and\ \bibinfo
  {author} {\bibfnamefont {K.}~\bibnamefont {Fukagata}},\ }\bibfield  {title}
  {\enquote {\bibinfo {title} {Model order reduction with neural networks:
  Application to laminar and turbulent flows},}\ }\href@noop {} {\bibfield
  {journal} {\bibinfo  {journal} {\rm{arXiv:2011.10277}\hspace{-0.3em}}\ }
  (\bibinfo {year} {2020})}\BibitemShut {NoStop}%
\bibitem [{\citenamefont {Morimoto}\ \emph {et~al.}(2020)\citenamefont
  {Morimoto}, \citenamefont {Fukami}, \citenamefont {Zhang},\ and\
  \citenamefont {Fukagata}}]{MFZF2020}%
  \BibitemOpen
  \bibfield  {author} {\bibinfo {author} {\bibfnamefont {M.}~\bibnamefont
  {Morimoto}}, \bibinfo {author} {\bibfnamefont {K.}~\bibnamefont {Fukami}},
  \bibinfo {author} {\bibfnamefont {K.}~\bibnamefont {Zhang}},\ and\ \bibinfo
  {author} {\bibfnamefont {K.}~\bibnamefont {Fukagata}},\ }\bibfield  {title}
  {\enquote {\bibinfo {title} {Generalization techniques of neural networks for
  fluid flow estimation},}\ }\href@noop {} {\bibfield  {journal} {\bibinfo
  {journal} {\rm{arXiv:2011.11911}\hspace{-0.3em}}\ } (\bibinfo {year}
  {2020})}\BibitemShut {NoStop}%
\bibitem [{\citenamefont {Nair}\ and\ \citenamefont {Hinton}(2010)}]{NH2010}%
  \BibitemOpen
  \bibfield  {author} {\bibinfo {author} {\bibfnamefont {V.}~\bibnamefont
  {Nair}}\ and\ \bibinfo {author} {\bibfnamefont {G.~E.}\ \bibnamefont
  {Hinton}},\ }\bibfield  {title} {\enquote {\bibinfo {title} {Rectified linear
  units improve restricted boltzmann machines},}\ }\href@noop {} {\bibfield
  {journal} {\bibinfo  {journal} {In Proc. 27th International Conference on
  Machine Learning}\ } (\bibinfo {year} {2010})}\BibitemShut {NoStop}%
\bibitem [{\citenamefont {Graves}, \citenamefont {Jaitly},\ and\ \citenamefont
  {Mohamed}(2013)}]{graves2013hybrid}%
  \BibitemOpen
  \bibfield  {author} {\bibinfo {author} {\bibfnamefont {A.}~\bibnamefont
  {Graves}}, \bibinfo {author} {\bibfnamefont {N.}~\bibnamefont {Jaitly}},\
  and\ \bibinfo {author} {\bibfnamefont {A.}~\bibnamefont {Mohamed}},\
  }\bibfield  {title} {\enquote {\bibinfo {title} {Hybrid speech recognition
  with deep bidirectional lstm},}\ }in\ \href@noop {} {\emph {\bibinfo
  {booktitle} {2013 IEEE workshop on automatic speech recognition and
  understanding}}}\ (\bibinfo {organization} {IEEE},\ \bibinfo {year} {2013})\
  pp.\ \bibinfo {pages} {273--278}\BibitemShut {NoStop}%
\bibitem [{\citenamefont {Bergstra}, \citenamefont {Yamins},\ and\
  \citenamefont {Cox}(2013)}]{BYC2013}%
  \BibitemOpen
  \bibfield  {author} {\bibinfo {author} {\bibfnamefont {J.}~\bibnamefont
  {Bergstra}}, \bibinfo {author} {\bibfnamefont {D.}~\bibnamefont {Yamins}},\
  and\ \bibinfo {author} {\bibfnamefont {D.}~\bibnamefont {Cox}},\ }\bibfield
  {title} {\enquote {\bibinfo {title} {Making a science of model search:
  Hyperparameter optimization in hundreds of dimensions for vision
  architectures},}\ }in\ \href@noop {} {\emph {\bibinfo {booktitle}
  {International conference on machine learning}}}\ (\bibinfo {organization}
  {PMLR},\ \bibinfo {year} {2013})\ pp.\ \bibinfo {pages}
  {115--123}\BibitemShut {NoStop}%
\bibitem [{\citenamefont {Maulik}\ \emph
  {et~al.}(2020{\natexlab{b}})\citenamefont {Maulik}, \citenamefont {Mohan},
  \citenamefont {Lusch}, \citenamefont {Madireddy}, \citenamefont
  {Balaprakash},\ and\ \citenamefont {Livescu}}]{MMLMBL2019}%
  \BibitemOpen
  \bibfield  {author} {\bibinfo {author} {\bibfnamefont {R.}~\bibnamefont
  {Maulik}}, \bibinfo {author} {\bibfnamefont {A.}~\bibnamefont {Mohan}},
  \bibinfo {author} {\bibfnamefont {B.}~\bibnamefont {Lusch}}, \bibinfo
  {author} {\bibfnamefont {S.}~\bibnamefont {Madireddy}}, \bibinfo {author}
  {\bibfnamefont {P.}~\bibnamefont {Balaprakash}},\ and\ \bibinfo {author}
  {\bibfnamefont {D.}~\bibnamefont {Livescu}},\ }\bibfield  {title} {\enquote
  {\bibinfo {title} {Time-series learning of latent-space dynamics for
  reduced-order model closure},}\ }\href@noop {} {\bibfield  {journal}
  {\bibinfo  {journal} {Physica D: Nonlinear Phenomena}\ }\textbf {\bibinfo
  {volume} {405}},\ \bibinfo {pages} {132368} (\bibinfo {year}
  {2020}{\natexlab{b}})}\BibitemShut {NoStop}%
\bibitem [{\citenamefont {Kingma}\ and\ \citenamefont {Ba}(2014)}]{kingma2014}%
  \BibitemOpen
  \bibfield  {author} {\bibinfo {author} {\bibfnamefont {D.~P.}\ \bibnamefont
  {Kingma}}\ and\ \bibinfo {author} {\bibfnamefont {J.}~\bibnamefont {Ba}},\
  }\bibfield  {title} {\enquote {\bibinfo {title} {Adam: A method for
  stochastic optimization},}\ }\href@noop {} {\bibfield  {journal} {\bibinfo
  {journal} {{\rm arXiv:1412.6980}\hspace{-0.3em}}\ } (\bibinfo {year}
  {2014})}\BibitemShut {NoStop}%
\bibitem [{\citenamefont {Prechelt}(1998)}]{prechelt1998}%
  \BibitemOpen
  \bibfield  {author} {\bibinfo {author} {\bibfnamefont {L.}~\bibnamefont
  {Prechelt}},\ }\bibfield  {title} {\enquote {\bibinfo {title} {Automatic
  early stopping using cross validation: quantifying the criteria},}\
  }\href@noop {} {\bibfield  {journal} {\bibinfo  {journal} {Neural Netw.}\
  }\textbf {\bibinfo {volume} {11}},\ \bibinfo {pages} {761--767} (\bibinfo
  {year} {1998})}\BibitemShut {NoStop}%
\bibitem [{\citenamefont {John}, \citenamefont {Parviz},\ and\ \citenamefont
  {Robert}(1987)}]{KMM1987}%
  \BibitemOpen
  \bibfield  {author} {\bibinfo {author} {\bibfnamefont {K.}~\bibnamefont
  {John}}, \bibinfo {author} {\bibfnamefont {M.}~\bibnamefont {Parviz}},\ and\
  \bibinfo {author} {\bibfnamefont {M.}~\bibnamefont {Robert}},\ }\bibfield
  {title} {\enquote {\bibinfo {title} {Turbulence statistics in fully developed
  channel flow at low reynolds number},}\ }\href@noop {} {\bibfield  {journal}
  {\bibinfo  {journal} {J. Fluid Mech.}\ }\textbf {\bibinfo {volume} {177}},\
  \bibinfo {pages} {133--166} (\bibinfo {year} {1987})}\BibitemShut {NoStop}%
\bibitem [{\citenamefont {Fukami}, \citenamefont {Fukagata},\ and\
  \citenamefont {Taira}(2021)}]{FFT2020b}%
  \BibitemOpen
  \bibfield  {author} {\bibinfo {author} {\bibfnamefont {K.}~\bibnamefont
  {Fukami}}, \bibinfo {author} {\bibfnamefont {K.}~\bibnamefont {Fukagata}},\
  and\ \bibinfo {author} {\bibfnamefont {K.}~\bibnamefont {Taira}},\ }\bibfield
   {title} {\enquote {\bibinfo {title} {Machine-learning-based spatio-temporal
  super resolution reconstruction of turbulent flows},}\ }\href@noop {}
  {\bibfield  {journal} {\bibinfo  {journal} {J. Fluid Mech.}\ }\textbf
  {\bibinfo {volume} {909}} (\bibinfo {year} {2021})}\BibitemShut {NoStop}%
\bibitem [{\citenamefont {Lee}\ and\ \citenamefont {You}(2019)}]{LY2019}%
  \BibitemOpen
  \bibfield  {author} {\bibinfo {author} {\bibfnamefont {S.}~\bibnamefont
  {Lee}}\ and\ \bibinfo {author} {\bibfnamefont {D.}~\bibnamefont {You}},\
  }\bibfield  {title} {\enquote {\bibinfo {title} {Data-driven prediction of
  unsteady flow fields over a circular cylinder using deep learning},}\
  }\href@noop {} {\bibfield  {journal} {\bibinfo  {journal} {J. Fluid Mech.}\
  }\textbf {\bibinfo {volume} {879}},\ \bibinfo {pages} {217--254} (\bibinfo
  {year} {2019})}\BibitemShut {NoStop}%
\bibitem [{\citenamefont {Raissi}, \citenamefont {Yazdani},\ and\ \citenamefont
  {Karniadakis}(2020)}]{raissi2020hidden}%
  \BibitemOpen
  \bibfield  {author} {\bibinfo {author} {\bibfnamefont {M.}~\bibnamefont
  {Raissi}}, \bibinfo {author} {\bibfnamefont {A.}~\bibnamefont {Yazdani}},\
  and\ \bibinfo {author} {\bibfnamefont {G.~E.}\ \bibnamefont {Karniadakis}},\
  }\bibfield  {title} {\enquote {\bibinfo {title} {Hidden fluid mechanics:
  Learning velocity and pressure fields from flow visualizations},}\
  }\href@noop {} {\bibfield  {journal} {\bibinfo  {journal} {Science}\ }\textbf
  {\bibinfo {volume} {367}},\ \bibinfo {pages} {1026--1030} (\bibinfo {year}
  {2020})}\BibitemShut {NoStop}%
\bibitem [{\citenamefont {Buzzicotti}\ \emph {et~al.}(2020)\citenamefont
  {Buzzicotti}, \citenamefont {Bonaccorso}, \citenamefont {Di~Leoni},\ and\
  \citenamefont {Biferale}}]{buzzicotti2020reconstruction}%
  \BibitemOpen
  \bibfield  {author} {\bibinfo {author} {\bibfnamefont {M.}~\bibnamefont
  {Buzzicotti}}, \bibinfo {author} {\bibfnamefont {F.}~\bibnamefont
  {Bonaccorso}}, \bibinfo {author} {\bibfnamefont {P.~C.}\ \bibnamefont
  {Di~Leoni}},\ and\ \bibinfo {author} {\bibfnamefont {L.}~\bibnamefont
  {Biferale}},\ }\bibfield  {title} {\enquote {\bibinfo {title} {Reconstruction
  of turbulent data with deep generative models for semantic inpainting from
  turb-rot database},}\ }\href@noop {} {\bibfield  {journal} {\bibinfo
  {journal} {{\rm arXiv:2006.09179}}\ } (\bibinfo {year} {2020})}\BibitemShut
  {NoStop}%
\bibitem [{\citenamefont {Yamamoto}\ and\ \citenamefont
  {Tsuji}(2018)}]{yamamoto2018numerical}%
  \BibitemOpen
  \bibfield  {author} {\bibinfo {author} {\bibfnamefont {Y.}~\bibnamefont
  {Yamamoto}}\ and\ \bibinfo {author} {\bibfnamefont {Y.}~\bibnamefont
  {Tsuji}},\ }\bibfield  {title} {\enquote {\bibinfo {title} {Numerical
  evidence of logarithmic regions in channel flow at {$Re_\tau=8000$}},}\
  }\href@noop {} {\bibfield  {journal} {\bibinfo  {journal} {Phys. Rev.
  Fluids}\ }\textbf {\bibinfo {volume} {3}},\ \bibinfo {pages} {012602}
  (\bibinfo {year} {2018})}\BibitemShut {NoStop}%
\bibitem [{\citenamefont {Kim}\ and\ \citenamefont
  {Lee}(2020{\natexlab{b}})}]{kim2020deep}%
  \BibitemOpen
  \bibfield  {author} {\bibinfo {author} {\bibfnamefont {J.}~\bibnamefont
  {Kim}}\ and\ \bibinfo {author} {\bibfnamefont {C.}~\bibnamefont {Lee}},\
  }\bibfield  {title} {\enquote {\bibinfo {title} {Deep unsupervised learning
  of turbulence for inflow generation at various reynolds numbers},}\
  }\href@noop {} {\bibfield  {journal} {\bibinfo  {journal} {J. Comput. Phys.}\
  }\textbf {\bibinfo {volume} {406}},\ \bibinfo {pages} {109216} (\bibinfo
  {year} {2020}{\natexlab{b}})}\BibitemShut {NoStop}%
\bibitem [{\citenamefont {Kim}\ \emph {et~al.}(2021)\citenamefont {Kim},
  \citenamefont {Kim}, \citenamefont {Won},\ and\ \citenamefont
  {Lee}}]{kim2020unsupervised}%
  \BibitemOpen
  \bibfield  {author} {\bibinfo {author} {\bibfnamefont {H.}~\bibnamefont
  {Kim}}, \bibinfo {author} {\bibfnamefont {J.}~\bibnamefont {Kim}}, \bibinfo
  {author} {\bibfnamefont {S.}~\bibnamefont {Won}},\ and\ \bibinfo {author}
  {\bibfnamefont {C.}~\bibnamefont {Lee}},\ }\bibfield  {title} {\enquote
  {\bibinfo {title} {Unsupervised deep learning for super-resolution
  reconstruction of turbulence},}\ }\href@noop {} {\bibfield  {journal}
  {\bibinfo  {journal} {J. Fluid Mech.}\ }\textbf {\bibinfo {volume} {910}},\
  \bibinfo {pages} {A29} (\bibinfo {year} {2021})}\BibitemShut {NoStop}%
\bibitem [{\citenamefont {Fukami}\ \emph {et~al.}(2021)\citenamefont {Fukami},
  \citenamefont {Maulik}, \citenamefont {Ramachandra}, \citenamefont
  {Fukagata},\ and\ \citenamefont {Taira}}]{FukamiVoronoi}%
  \BibitemOpen
  \bibfield  {author} {\bibinfo {author} {\bibfnamefont {K.}~\bibnamefont
  {Fukami}}, \bibinfo {author} {\bibfnamefont {R.}~\bibnamefont {Maulik}},
  \bibinfo {author} {\bibfnamefont {N.}~\bibnamefont {Ramachandra}}, \bibinfo
  {author} {\bibfnamefont {K.}~\bibnamefont {Fukagata}},\ and\ \bibinfo
  {author} {\bibfnamefont {K.}~\bibnamefont {Taira}},\ }\bibfield  {title}
  {\enquote {\bibinfo {title} {Global field reconstruction from sparse sensors
  with voronoi tessellation-assisted deep learning},}\ }\href@noop {}
  {\bibfield  {journal} {\bibinfo  {journal} {arXiv:2101.00554}\ } (\bibinfo
  {year} {2021})}\BibitemShut {NoStop}%
\bibitem [{\citenamefont {Kashefi}, \citenamefont {Rempe},\ and\ \citenamefont
  {Guibas}(2020)}]{kashefi2020point}%
  \BibitemOpen
  \bibfield  {author} {\bibinfo {author} {\bibfnamefont {A.}~\bibnamefont
  {Kashefi}}, \bibinfo {author} {\bibfnamefont {D.}~\bibnamefont {Rempe}},\
  and\ \bibinfo {author} {\bibfnamefont {L.~J.}\ \bibnamefont {Guibas}},\
  }\bibfield  {title} {\enquote {\bibinfo {title} {A point-cloud deep learning
  framework for prediction of fluid flow fields on irregular geometries},}\
  }\href@noop {} {\bibfield  {journal} {\bibinfo  {journal} {arXiv:2010.09469}\
  } (\bibinfo {year} {2020})}\BibitemShut {NoStop}%
\bibitem [{\citenamefont {Ogoke}\ \emph {et~al.}(2020)\citenamefont {Ogoke},
  \citenamefont {Meidani}, \citenamefont {Hashemi},\ and\ \citenamefont
  {Farimani}}]{ogoke2020graph}%
  \BibitemOpen
  \bibfield  {author} {\bibinfo {author} {\bibfnamefont {F.}~\bibnamefont
  {Ogoke}}, \bibinfo {author} {\bibfnamefont {K.}~\bibnamefont {Meidani}},
  \bibinfo {author} {\bibfnamefont {A.}~\bibnamefont {Hashemi}},\ and\ \bibinfo
  {author} {\bibfnamefont {A.~B.}\ \bibnamefont {Farimani}},\ }\bibfield
  {title} {\enquote {\bibinfo {title} {Graph convolutional neural networks for
  body force prediction},}\ }\href@noop {} {\bibfield  {journal} {\bibinfo
  {journal} {arXiv:2012.02232}\ } (\bibinfo {year} {2020})}\BibitemShut
  {NoStop}%
\bibitem [{\citenamefont {Tencer}\ and\ \citenamefont
  {Potter}(2020)}]{tencer2020enabling}%
  \BibitemOpen
  \bibfield  {author} {\bibinfo {author} {\bibfnamefont {J.}~\bibnamefont
  {Tencer}}\ and\ \bibinfo {author} {\bibfnamefont {K.}~\bibnamefont
  {Potter}},\ }\bibfield  {title} {\enquote {\bibinfo {title} {Enabling
  nonlinear manifold projection reduced-order models by extending convolutional
  neural networks to unstructured data},}\ }\href@noop {} {\bibfield  {journal}
  {\bibinfo  {journal} {arXiv:2006.06154}\ } (\bibinfo {year}
  {2020})}\BibitemShut {NoStop}%
\bibitem [{\citenamefont {Morimoto}\ \emph {et~al.}(2021)\citenamefont
  {Morimoto}, \citenamefont {Fukami}, \citenamefont {Zhang}, \citenamefont
  {Nair},\ and\ \citenamefont {Fukagata}}]{morimoto2101convolutional}%
  \BibitemOpen
  \bibfield  {author} {\bibinfo {author} {\bibfnamefont {M.}~\bibnamefont
  {Morimoto}}, \bibinfo {author} {\bibfnamefont {K.}~\bibnamefont {Fukami}},
  \bibinfo {author} {\bibfnamefont {K.}~\bibnamefont {Zhang}}, \bibinfo
  {author} {\bibfnamefont {A.~G.}\ \bibnamefont {Nair}},\ and\ \bibinfo
  {author} {\bibfnamefont {K.}~\bibnamefont {Fukagata}},\ }\bibfield  {title}
  {\enquote {\bibinfo {title} {Convolutional neural networks for fluid flow
  analysis: toward effective metamodeling and low-dimensionalization},}\
  }\href@noop {} {\bibfield  {journal} {\bibinfo  {journal} {arXiv:2101.02535}\
  } (\bibinfo {year} {2021})}\BibitemShut {NoStop}%
\bibitem [{\citenamefont {Liu}, \citenamefont {Kutz},\ and\ \citenamefont
  {Brunton}(2019)}]{LKB2020}%
  \BibitemOpen
  \bibfield  {author} {\bibinfo {author} {\bibfnamefont {Y.}~\bibnamefont
  {Liu}}, \bibinfo {author} {\bibfnamefont {J.~N.}\ \bibnamefont {Kutz}},\ and\
  \bibinfo {author} {\bibfnamefont {S.~L.}\ \bibnamefont {Brunton}},\
  }\bibfield  {title} {\enquote {\bibinfo {title} {Hierarchical deep learning
  of multiscale differential equation time-steppers},}\ }\href@noop {}
  {\bibfield  {journal} {\bibinfo  {journal} {{\rm arXiv:2008.09768}}\ }
  (\bibinfo {year} {2019})}\BibitemShut {NoStop}%
\end{thebibliography}%
\end{document}